\documentclass[journal]{IEEEtran}

 \usepackage[pdftex]{graphicx}
 \usepackage[cmex10]{amsmath}
 \usepackage{amssymb}
 \usepackage{xcolor}
\usepackage{amsthm}
\usepackage{algorithmic}

\allowdisplaybreaks[4]

\theoremstyle{plain}
\newtheorem{theorem}{Theorem}[section]

\theoremstyle{definition}

\theoremstyle{remark}

\title{A universal, operational theory of unicast multi-user communication with fidelity criteria}
\author{Mukul Agarwal,
        Sanjoy Mitter,~\IEEEmembership{Fellow,~IEEE}
        and Anant Sahai,~\IEEEmembership{Member,~IEEE}}

\begin{document}

\maketitle

\begin{abstract}

This is a three part paper.

Optimality of source-channel separation for communication with a fidelity criterion when the channel is compound as defined in \cite{CsiszarKorner} and general as defined in \cite{VerduHan} is proved in Part I. It is assumed that random codes are permitted. The word ``universal'' in the title of this paper refers to the fact that the channel model is compound. The proof uses a layered black-box or a layered input-output view-point. In particular, only the end-to-end description of the channel as being capable of communicating a source to within a certain distortion level is used when proving separation. This implies that the channel model does not play any role for separation to hold as long as there is a source model. Further implications of the layered black-box view-point are discussed.

Optimality of source-medium separation  for multi-user communication with  fidelity criteria over a general, compound medium in the unicast setting is proved in Part II, thus generalizing Part I to the unicast, multi-user setting.

Part III gets to an understanding of the question, ``\emph{Why} is a channel which is capable of communicating a source to within a certain distortion level, also capable of communicating bits at any rate less than the infimum of the rates needed to code the source to within the distortion level'': this lies at the heart of why optimality of separation for communication with a fidelity criterion holds. The perspective taken to get to this understanding is a randomized covering-packing perspective, and the proof is operational.

\end{abstract}

\section{Part I: point-to-point setting}

\subsection{Introduction to, and contribution of Part I} \label{Introduction}

Optimality of separation based architectures for communication with a fidelity criterion over a discrete memoryless channel is proved in \cite{Shannon}. Optimality refers to the fact that if communication of some source to within some distortion level can be accomplished with some architecture, communication of the same source to within the same distortion level can also be accomplished with a source-channel separation based architecture. This result can be generalized to indecomposable channels, that is, finite state Markoff channels for which the state information dies down with time, as defined rigorously in \cite{GallagerInformationTheory}. Part I generalizes this optimality to the case when the channel is compound, that is, the channel belongs to a set, as defined in \cite{CsiszarKorner}, and the channel transition probability is general, as defined in \cite{VerduHan}, with the difference that the probability of excess distortion criterion which is essentially the same as the criterion in \cite{CsiszarKorner} is used as the fidelity criterion instead of the expected distortion criterion used in \cite{Shannon}. The use of the word ``universal'' in the title of this paper refers to the fact that the result holds for a compound channel. Note that the universality is over the channel, not the source.

A generalization to the compound setting is needed because the action of real media like wireless medium or the internet cannot be modeled as a known transition probability and one way to model these media might be that they belong to a set of transition probabilities. The multi-user generalization of Part I in the unicast setting is the subject of Part II.

In order to prove this optimality, encoders and decoders are allowed to be random. That is, the encoder is a probability distribution on the set of deterministic encoders, the decoder has access to the particular realization of the encoder, and based on this access, acts as a probability distribution on the set of deterministic decoders. Error is calculated by averaging over the random code. This is called random coding. Random-coding argument in \cite{ShannonReliable} uses random codes. The difference is that in \cite{ShannonReliable}, random-coding is a proof technique whereas in the argument in Part I, random-coding is necessary in the sense that separation is not optimal for communication with a fidelity criterion over a general, compound channel if random codes are not permitted.

The proof uses a layered black-box or a layered input-output view-point, and the proof style is different from that used in  \cite{Shannon}. The question arises: can a  proof in the style similar to \cite{Shannon} be used to prove the optimality of separation for communication with a fidelity criterion when the channel is compound. The answer is yes: Amos Lapidoth showed the first author, how to prove the result using techniques similar to \cite{Shannon} in a private discussion when the first author was explaining the result to Amos Lapidoth. A further question arises: what is the need for a different proof technique? The answer is many fold:  First, the proof technique in \cite{Shannon} and the further extension due to Amos Lapidoth require an indecomposability assumption on the channel whereas the proof technique in Part I works for general channels as defined in \cite{VerduHan}. It is for this reason that the probability of excess distortion criterion is used instead of the expected distortion criterion. The use of the probability of excess distortion criterion instead of the expected distortion criterion is similar in spirit as the use of the $\inf$ information rate in \cite{VerduHan} instead of mutual information: in \cite{VerduHan}, a formula for channel capacity is given in terms of the $\inf$ information rate which is the  $\lim \inf$ in probability (see \cite{VerduHan}) of the normalized information densities where the information density is
\begin{align}
i_{X^nW^n}(a^n; b^n) \triangleq \log \frac{P_{Y^n|X^n}(b^n|a^n)}{P_{Y^n}(b^n)}
\end{align}
as compared to the usual formula for channel capacity which is given in terms of mutual information which is the expected value of the normalized information density; similarly, the probability of excess distortion criterion is a criterion in terms of a slight variant of the $\limsup$ in probability (see \cite{VerduHan}) of $\displaystyle{\frac{1}{n}d^n(X^n, Y^n)}$ whereas the expected distortion criterion is a criterion in terms of the expectation of $\displaystyle{\frac{1}{n}d^n(X^n, Y^n)}$. The result is similar in spirit too: optimality of separation for communication with a fidelity criterion can be proved for a general channel and correspondingly, the formula for channel capacity in \cite{VerduHan} holds for a general channel. Second, the layered black-box view-point uses only the end-to-end description of the channel as being capable of communicating a source to within a certain distortion level to prove separation. This implies that the channel model does not play any role for separation to hold as long as there is a source model. This implication cannot be derived from a proof in the style of \cite{Shannon}. This implication that the channel model does not play any role is also true in \cite{LomnitzFeder} where the context is not proving the optimality of separation but reliable communication of bits over an individual channel. Further, the layered black-box view-point has architectural implications. These and other implications of the layered black-box view-point which do not follow from the Shannon-Lapidoth view-point will be discussed later in the paper.

Part I is joint work of all three authors. Part II and Part III are the joint work of the first author and the second author.

\subsection{Notation and definitions for Part I} \label{NotAndDef}

Superscript $n$ will denote a quantity related to  block-length is $n$. For example, $x^n$ will be the channel input when the block-length is $n$.  The only exception to this rule is for a real number: $\omega_n$ is used corresponding to block-length $n$ if $\omega_n$ is a real number in order not to confuse $\omega^n$ with the $n^{th}$ power of $\omega$. As the block-length varies, $x = <x^n>_1^\infty$ will denote the sequence for various block-lengths.

The channel input space is $\mathcal I$ and the channel output space is $\mathcal O$. $\mathcal I$ and $\mathcal O$ are finite sets. The channel is a sequence $k = <k^n>_1^\infty$ where 
\begin{align}
k^n &: \mathcal I^n  \rightarrow \mathcal P(\mathcal O^n) \nonumber \\
       &: \iota^n            \mapsto k^n(\cdot|\iota^n)
\end{align}
When the block-length is $n$, the channel acts as $k^n(\cdot  | \cdot)$; $k^n(o^n|\iota^n)$ is the probability that the channel output is $o^n$ if the channel input is $\iota^n$.  This model is the same as the model in \cite{VerduHan} which is the same as the model at the top of Page 100 of \cite{CsiszarKorner}. The model of a ``physical'' channel would impose causality and nestedness among the various $k^n$, and this would be a special case of the above model. A compound channel is a set $k \in \mathcal A$ of channels with input space $\mathcal I$ and output space $\mathcal O$. This model of a compound channel is the same as the model of a compound channel defined in \cite{CsiszarKorner} though the emphasis in \cite{CsiszarKorner} is on compound DMCs. This model of a compound channel also generalizes arbitrarily varying channels defined in \cite{CsiszarKorner}.

The source input space is $\mathcal X$ and the source reproduction space is $\mathcal Y$. $\mathcal X$ and $\mathcal Y$ are finite sets. $X$ is a random variable on $\mathcal X$. $X = <X^n>_1^\infty$ is the i.i.d. $X$ source where $X^n$ is i.i.d. $X$ sequence of length $n$.  $d: \mathcal X \times \mathcal Y \rightarrow [0, \infty )$ is a distortion function. The $n$-letter distortion function is defined additively:
\begin{align}\label{AdditiveDistortionFunction}
d^n(x^n, y^n) \triangleq \sum_{i=1}^n d(x^n(i), y^n(i))
\end{align}

Communication of the i.i.d. $X$ source over $k \in \mathcal A$ requires an encoder and decoder. When the block-length is $n$, a deterministic encoder is a map $e^n:\mathcal X^n \rightarrow \mathcal I^n$ and a deterministic decoder is a map $f^n: \mathcal O^n \rightarrow \mathcal Y^n$. 
Encoder and decoder can be random; a random encoder-decoder can be defined in two ways:
\begin{itemize}
\item
Encoder is a probability distribution on the space of deterministic encoders and the decoder, based on the knowledge of the encoder realization, acts as a transition probability from the set of deterministic encoders to the set of probability distributions on the set of deterministic decoders
\item
A joint probability distribution on the space of deterministic encoders and decoders
\end{itemize}
A random encoder-decoder can be realized if there is a shared continuous valued random variable independent of all other random variables in the system at both the encoder and the decoder. A precise discussion of a random encoder-decoder is omitted. The encoder-decoder sequence is denoted by $<e^n, f^n>_1^\infty$. If the input to the encoder $e^n$ is $X^n$, the composition $e^n \circ k^n \circ f^n$ produces an output $Y^n$ which might depend on the particular $k \in \mathcal A$. Channel $k \in \mathcal A$ is said to be capable of communicating (end-to-end) the i.i.d. $X$ source to within a distortion $D$ if there exist encoder-decoder $<e^n, f^n>_1^\infty$ independent of the particular $k \in \mathcal A$ and a non-negative real sequence $<\omega_n>_1^\infty$, $\displaystyle {\lim_{n \to \infty} \omega_n = 0}$ independent of the particular $k \in \mathcal A$ such that
\begin{align} \label{PEDCriterionChannel}
\Pr \left ( \frac{1}{n}d^n(X^n, Y^n) > D \right ) \leq \omega_n \  \forall k \in \mathcal A
\end{align}
$<\omega_n>_1^\infty$ causes a uniformity in the rate at which the probability of excess distortion tends to $0$ as block-length tends to $\infty$ over $k \in \mathcal A$.

In a separation-based architecture, the encoder is the composition of a source encoder and the channel encoder and the decoder is the composition of a channel decoder and a source decoder. The action of the source encoder-source decoder pair which is used depends on the source and not on the channel. The action of the channel encoder-channel decoder pair which is used depends on the channel and not on the source. In this sense, the source encoder-source decoder pair and the channel encoder-channel decoder pair act in a way to separate the source and the channel. Further, the communication over the channel with the help of channel encoder and channel decoder is reliable communication; fixing this notion of communication independent of the channel leads to architectural standardization in the sense that irrespective of the channel of communication, reliable communication is carried out over the channel as part of the end-to-end communication: this point is of importance in the multi-user setting; see Chapter 1 of \cite{GallagerDigital}. Also see Chapter 1 of \cite{GallagerDigital} for a more detailed high-level discussion of separation, and its importance. Mathematically, separation architectures are discussed below:

Let 
\begin{align}
\mathcal M^n_R \triangleq \{1, 2, \ldots, 2^{\lfloor nR \rfloor} \}
\end{align}
$\mathcal M^n_R$ is the message set. When the block-length is $n$, a rate $R$ deterministic source encoder  is $e_s^n: \mathcal X^n \rightarrow \mathcal M_R^n$ and a rate $R$ deterministic source decoder $f_s^n: \mathcal M_R^n \rightarrow \mathcal Y^n$.  $(e_s^n, f_s^n)$ is the block-length $n$ rate $R$  deterministic source-code. The source-code is allowed to be random in the sense discussed previously. $<e_s^n, f_s^n>_1^\infty$ is the rate $R$ source-code. An example of a source code used in information-theory arguments is one which generate codewords i.i.d. from a particular source: this is an example of a random source-code.
The source-code $<e_s^n, f_s^n>_1^\infty$ is said to code the i.i.d. $X$ source to within a distortion $D$ if with input $X^n$ to $e_s^n \circ f_s^n$, the output is $Y^n$ such that
\begin{align} \label{SourceCodeD}
\lim_{n \to \infty} \Pr \left ( \frac{1}{n} d^n(X^n, Y^n) > D \right ) = 0
\end{align}
The above criterion is the probability of excess distortion criterion and is \emph{essentially the same} as the criterion used in Chapter $2$, $\S 2$ of \cite{CsiszarKorner}. The infimum of rates needed to code the i.i.d. $X$ source to within the distortion $D$ when the distortion function is $d$ under the probability of excess distortion criterion is the rate-distortion function $R^P_X(D)$; note that the dependence of the rate-distortion function on $d$ is suppressed in the notation. The above is the operational definition of the rate-distortion function. Rate-distortion function can also be defined information-theoretically:
\begin{align}\label{JInformationTheoreticRateDistortionFunction}
R^I_X(D) \triangleq \inf_{ \left \{ p_{Y|X}\ : \   \sum_{x \in \mathcal X, y \in \mathcal Y} p_X(x)p_{Y|X}(y|x) \leq D \right \} } I(X;Y) 
\end{align}
An expression for $R^P_X(D)$ is $R^I_X(D)$. This is because $R^P_X(D)$  is equal the rate-distortion function $R(D)$ defined in Chapter $2$, $\S 2$ of \cite{CsiszarKorner}: this follows from the similarity in distortion criteria used to define $R^P_X(D)$ and $R(D)$, and it is proved in \cite{CsiszarKorner} that an expression for $R(D)$ is $R^I_X(D)$. 

When the block-length is $n$, a rate $R$ deterministic channel encoder is a map $e_c^n:\mathcal M_R^n \rightarrow \mathcal I^n$ and a rate $R$ deterministic channel decoder is a map $f_c^n: \mathcal O^n \rightarrow \hat {\mathcal M}_R^n$ where $\hat {\mathcal M}_R^n \triangleq \mathcal M_R^n \cup \{e\}$ is the message reproduction set where `e' denotes error. The encoder and decoder are allowed to be random in the sense discussed previously. $<e_c^n, f_c^n>_1^\infty$ is the rate $R$ channel code. The classic argument used in \cite{ShannonReliable} to derive the achievability of the mutual information expression for channel capacity uses a random code. Denote 
\begin{align}
g\in  \mathcal G_{\mathcal A} \triangleq \{<e_c^n \circ k^n \circ f_c^n>_1^\infty\ | \ k \in \mathcal A \}
\end{align}
$g \in \mathcal G_{\mathcal A}$ is a compound channel with input space $<\mathcal M_R^n>_1^\infty$ and output space $<\hat {\mathcal M}_R^n>_1^\infty$. Rate $R$ is said to be reliably achievable over $k \in \mathcal A$ if there exists a  rate $R$ channel code $<e_c^n, f_c^n>_1^\infty$ and a sequence $<\delta_n>_1^\infty$, $\delta_n \to 0$ as $n \to \infty$ such that
\begin{align}
 \sup_{m^n \in \mathcal M_R^n} g^n(\{ m^n\}^c|m^n) \leq \delta_n \ \forall k \in \mathcal A
\end{align}
$<\delta_n>_1^\infty$ causes a uniformity in the rate at which the maximal error probability tends to zero as block-length tends to $\infty$ and plays a role similar to the sequence $<\omega_n>_1^\infty$ in (\ref{PEDCriterionChannel}). Since random codes are permitted, if rate $R$ is achievable, so is any rate less than $R$. Supremum of all achievable rates is the capacity of $k \in \mathcal A$.

$<e_s^n \circ e_c^n, f_c^n \circ f_s^n>_1^\infty$ is the separation-based encoder-decoder. 

The question is: if there exists some architecture to communicate the i.i.d. $X$ source to within a distortion $D$ over $k \in \mathcal A$, does there also exist a separation architecture to communicate the i.i.d. $X$ source to within a distortion $D$ over $k \in \mathcal A$? Note that this is an end-to-end question regarding communication of the i.i.d. $X$ source over $k \in \mathcal A$ and not a question about just source-coding or channel-coding. Under certain assumptions, this question is answered in the affirmative in the next subsection.

\subsection{Optimality of separation for communication with a fidelity criterion over a general, compound channel}

\subsubsection{Theorem and proof}

\begin{theorem}[Optimality of separation] \label{KKUniXKK}
Assume that random codes are permitted. Let $k \in \mathcal A$ be capable of communicating the i.i.d. $X$ source to within a distortion level $D$ under an additive distortion function $d$. Then, reliable communication can be accomplished over $k$ at rates $<R^P_X(D)$. This reliable communication can be accomplished with consumption of channel resources which is the same as the channel resource consumption in the original architecture which communicates the i.i.d. $X$ source to within a distortion $D$ over  $k \in \mathcal A$. 

Further, if  reliable communication can be accomplished over $k \in \mathcal A$ at a certain rate strictly $>R^P_X(D)$, then the i.i.d. $X$ source can be communicated to within a distortion $D$ over $k \in \mathcal A$ by use of a separation architecture. The channel resource consumption in this separation architecture is the same as the channel resource consumption in the architecture for reliable communication at rate strictly $>R^P_X(D)$ when the distribution on the message set is uniform.
\end{theorem}

\begin{IEEEproof}
$k \in \mathcal A$ is capable of communicating the i.i.d. $X$ source to within a distortion $D$ with the help of some encoder-decoder $<e^n, f^n>_1^\infty$. Consider the channel set
\begin{align}
\mathcal C_{\mathcal A} \triangleq \{ <e^n \circ k^n \circ f^n>_1^\infty, k \in \mathcal A \}
\end{align}
$c = <c^n>_1^\infty \in \mathcal C_{\mathcal A}$ is a compound channel with input space $\mathcal X$ and output space $\mathcal Y$. It will be proved that by use of some encoder-decoder $<E^n, F^n>_1^\infty$, reliable communication can be accomplished over $c \in \mathcal C_{\mathcal A}$ at rates $<R^P_X(D)$ with consumption of same channel resources as the architecture $<e^n \circ k^n \circ f^n>_1^\infty$, when used for communicating the i.i.d. $X$ source to within a distortion $D$. It will follow that by use of encoder-decoder $<E^n \circ e^n, f^n \circ F^n>_1^\infty$, reliable communication can be accomplished over $k \in \mathcal A$ at rates $<R^P_X(D)$ with consumption of same channel resources as the architecture $<e^n \circ k^n \circ f^n>_1^\infty$, when used for communicating the i.i.d. $X$ source to within a distortion $D$. 

This will be done by use of a random-coding argument. Let the block-length be $n$.

\emph{Codebook generation:} Generate $2^{\lfloor nR\rfloor}$ codewords i.i.d. $X$. This is the codebook $\mathcal K^n$ which the encoder $E^{n}$ uses.

\emph{Joint typicality:} $(x^n, y^n)$, $x^n \in \mathcal X^n$, $y^n \in \mathcal Y^n$ are said to be $\epsilon$ jointly typical if
\begin{enumerate}
\item
$x^n$ is  $\epsilon$ $p_X$ typical, that is, ${x^n} \in \mathcal T(p_X, \epsilon)$
\item
$\displaystyle{\frac{1}{n}d^n(x^n, y^n) \leq D}$
\end{enumerate}

\emph{Decoding}: Let $y^n$ be received as the output of the channel. If $\exists$ unique $x^n \in$ the code book $\mathcal K^n$ such that $(x^n, y^n)$ $\epsilon$ jointly typical, declare that $x^n$ is transmitted, else declare error. This is the decoder $F^n$.

With this encoding-decoding procedure, it can be proved that reliable communication can be accomplished over $c \in \mathcal C_{\mathcal A}$ at rates $<R^P_X(D)$: the error analysis argument for this is in Appendix \ref{AppendixErrorAnalysis}. This proves that reliable communication can be accomplished over $k \in \mathcal A$ at rates $<R^P_X(D)$.

Next, the channel resource consumption required in this architecture for reliable communication is considered.The encoder $E^n$ generates i.i.d. $X$ codes. Thus, the input to $c^n = e^n \circ k^n \circ f^n$ is i.i.d. $X$ which is the same as the input $X^n$ to the original architecture $e^n \circ k^n \circ f^n$. It follows that the input to the channel $k^n$ in the architecture for reliable communication is $I^{s, n}$ which is the same \emph{in distribution} as the input $I^n$ to $k^n$ in the original architecture. It follows that the channel resource consumption in the architecture for reliable communication is the same as the channel resource consumption in the original architecture. Note that $I^n$ might depend on the particular $k \in \mathcal A$ but this is irrelevant to the argument. 

This proves the first part of the theorem.

The proof of the second part of the theorem is the usual argument of source coding followed by channel coding. Briefly, the argument is the following:  let rate $R^P_X(D) + \epsilon$ be reliably achievable over $k \in \mathcal A$. Encode the i.i.d. $X$ source to within a distortion $D$ at rate $R^P_X(D) +  \epsilon$. Communicate this rate $R^P_X(D) + \epsilon$ message reliably over $k \in \mathcal A$. Then, source-decode the message reproduction. End-to-end, the i.i.d. $X$ source is communicated to within a distortion $D$ over $k \in \mathcal A$. A detailed argument is omitted. Since random codes are permitted, by using a symmetrically permuted codebook, the distribution on the output of the source encoder can be made uniform. It follows by an argument similar to the resource consumption argument before that the channel resource consumption in the separation architecture is the same as the channel resource consumption in the architecture for reliable communication at rate $R^P_X(D) + \epsilon$ when the distribution on the message set is uniform. A full argument is omitted.

This finishes the proof of the theorem.
\end{IEEEproof}

Random codes are essential: an example of the failure of the optimality of separation for communication with a fidelity criterion when the channel is general but random codes are not permitted is in \cite{MukulSwastikMitter}. The authors conjecture that random codes are not needed if $\mathcal A$ is compact.

\subsubsection{Extensions}

The source and channel have been assumed to evolve on the same time scale in Theorem \ref{KKUniXKK}. This assumption has been made only for mathematical convenience and can be removed.

Theorem \ref{KKUniXKK} can be generalized to continuous valued alphabet: discretize the continuous valued space, make it discrete and then take a limit as the discretization size $\to 0$; some details are in \cite{AgarwalSahaiMitterAllerton}. 

The authors conjecture that Theorem \ref{KKUniXKK} can be generalized to ``many'' stationary ergodic sources.

The authors conjecture that Theorem \ref{KKUniXKK} can be generalized to continuous time source and channel evolution. Generalization to continuous time channel evolution: what matters in the proof of Theorem \ref{KKUniXKK} is the end-to-end description that the i.i.d. $X$ source is communicated to within a distortion $D$ and whether the channel evolves in continuous time or discrete time is immaterial. The usual approach to handle sources evolving in continuous time is to assume the source is band limited and use sampling theorem to make the source evolve in discrete time. This approach does not work because an additive distortion function in continuous time (this would be an integral over time instead of summation over discrete time as defined in (\ref{AdditiveDistortionFunction})) does not remain additive after sampling. The arguments would need to be carried out either in the continuous time domain or by discretizing time and taking a limit as the time discretization $\to 0$. Ideas from \cite{GrayPursley} might be helpful. 

The authors conjecture that universality in Theorem \ref{KKUniXKK} can be extended to the source; ideas from \cite{Ziv} and \cite{Ziv2} might be helpful.

\subsection{A note on, and implications of, the layered black-box view-point} \label{LBBImplications}
 
The proof of Theorem \ref{KKUniXKK} uses a layered black-box view-point:
\begin{itemize}
\item
The proof is layered in nature because encoder-decoder  $<E^n, F^n>_1^\infty$ is layered on top of the original architecture $<e^n \circ k^n \circ f^n>_1^\infty$
\item
The proof is black-box in nature because it uses only the end-to-end description of $<e^n \circ k^n \circ f^n>_1^\infty$ as communicating the i.i.d. $X$ source to within a distortion $D$.
\end{itemize}
Thus, a relationship is seen between two major constructs of information theory, namely channel capacity and the rate-distortion function, and how they are related in the black-box sense. 

The first part of Theorem \ref{KKUniXKK} is the converse. This converse is proved in \cite{Shannon} when the channel is a DMC. The proof uses definitions and properties involving entropy and mutual information. Such proof techniques are generally used for proving converse results in information theory. The proof in Part I uses a random-coding argument and thus proves a converse using an achievability technique. This is interesting in its own right.

The question arises: can a proof similar to \cite{Shannon} be given to prove the first part of Theorem \ref{KKUniXKK} for a compound DMC. The answer is yes and this proof is due to Amos Lapidoth; this proof was showed by Amos Lapidoth to the first author in a private discussion when the first author was  communicating Theorem \ref{KKUniXKK} to Amos Lapidoth. This proof is in Appendix \ref{ShannonLapidothView} as written by the first author, and has been included with the permission of Amos Lapidoth. In a nut-shell, the proof is: Use the expected distortion criterion instead of the probability of excess distortion criterion and denote the rate-distortion function under the expected distortion criterion by $R^E_X(D)$. Define 
\begin{align}
C^I (k \in \mathcal A) \triangleq \sup_{Q \in \mathcal P (\mathcal I)} \inf _ {k \in \mathcal A} I(Q, k)
\end{align}
where $k$ is now a DMC and $k$ in the above expression denotes the transition probability corresponding to the DMC. Let $k$ be capable of communicating the i.i.d. $X$ source to within a distortion $D$ under the expected distortion criterion, that is, there exist encoder-decoder $<e^n, f^n>_1^\infty$ independent of the particular $k \in \mathcal A$ and $<\omega_n>_1^\infty$, $\displaystyle{\lim_{n \to \infty} \omega_n = 0}$ independent of the particular $k \in \mathcal A$ such that with input $<X^n>_1^\infty$ to $<e^n \circ k^n \circ f^n>_1^\infty$, the output is $<Y^n>_1^\infty$ such that 
\begin{align}
E \left [ \frac{1}{n} d^n(X^n, Y^n) \right ] \leq D + \omega_n \ \forall k \in \mathcal A
\end{align}
Prove that $C^I(k \in \mathcal A) \geq R^I_X(D)$ by use of definition and computations involving entropy and mutual information: this proof closely parallels the proof in \cite{Shannon}. From \cite{CsiszarKorner}, capacity of $k \in \mathcal A$ is $C^I(k \in \mathcal A)$. From \cite{Shannon},  $R^E_X(D) = R^I_X(D)$. Thus, capacity of $k \in \mathcal A$ is $\geq R^E_X(D)$. This finishes the proof. This proof does not use layering on top of the existing architecture, nor does it use a black-box perspective. In what follows, this proof/view will be called the Shannon-Lapidoth proof/view.

A further question arises: what is the need of a different proof technique if the old proof technique can be used to prove the result. The rest of this subsection discusses various points which come out of the layered black-box proof and which are interesting in their own right, and further discusses the fact that they do not come out of the Shannon-Lapidoth proof.

The layered black-box proof holds for a general channel. The Shannon-Lapidoth proof can be generalized to indecomposable channels but some ergodicity assumption is required: an example of a non-indecomposable channel for which separation for communication with a fidelity criterion fails to hold when the expected distortion criterion is used can be found on Page $1$ of \cite{VerduHan}, ``Consider a binary channel where the output codeword is equal to the transmitted codeword with probability $\frac{1}{2}$ and independent of the transmitted codeword with probability $\frac{1}{2}$'': if the distortion is hamming distortion, and the input to this channel is a rate 1 bit stream, the expected distortion is $0.25$, $R^E_X(0.25) > 0$, but capacity of this channel is zero, and this implies failure of optimality of separation. 

The layered black-box perspective implies that for the optimality of separation for communication with a fidelity criterion to hold, the channel model does not play any role as long as there is a source model : the only requirement is that $<e^n \circ k^n \circ f^n>_1^\infty$ communicates the i.i.d. $X$ source to within a distortion $D$. This does not follow from the Shannon-Lapidoth proof: the channel is assumed to be a DMC or indecomposable. 

The layered black-box proof is \emph{semi-}operational.  It is operational in the sense that it uses only the operational meaning of channel capacity as the maximum rate of reliable communication. However, it is only semi-operational because it uses the definition of the information-theoretic rate-distortion function $R^I_X(D)$ and the equality of the operational rate-distortion function $R^P_X(D)$ and $R^I_X(D)$ crucially. The Shannon-Lapidoth proof is \emph{not} operational: it uses the information-theoretic channel capacity $C^I(k \in \mathcal A)$ and the information-theoretic rate-distortion function $R^I_X(D)$ and their equality with operational channel capacity of $k \in \mathcal A$ and the operational rate-distortion function $R^E_X(D)$, respectively, crucially. 

The layered black-box view-point gives insights into layered architectures for communication. Consider a system $s$ which communicates the i.i.d. $X$ source to within a distortion $D$ under the probability of excess distortion criterion. Let $X'$ be an i.i.d. source such that $R^P_{X'}(D') < R^P_X(D)$. Then, the i.i.d. $X'$ source can be communicated over the system $s$ to within a distortion $D'$ by layering, and thus, the system $s$ does not need to be ``broken''. The rough proof is: code the source $X'$ to within a distortion $D'$. The output is a rate $R^P_{X'}(D')$ bit stream. This bit stream can be communicated over $s$ by layering on top by the first part of Theorem \ref{KKUniXKK} and its proof since $R^P_{X'}(D') < R^P_X(D)$. Finally, decode the source. This layered architecture does not come out of the Shannon-Lapidoth view-point: the only conclusion that can be drawn is that the i.i.d. $X'$ source can be communicated to within a distortion $D'$ under the expected distortion criterion over an indecomposable sub-system of $s$.

A further application of the layered black-box view-point is in Part III: Part III answers the question, ``\emph{Why} is a channel which is capable of communicating a source to within a certain distortion level, capable of communicating bits at any rate less than the infimum of the rates needed to code the source to within the distortion level,'' by use of a layered black-box, randomized covering-packing perspective. This question lies at the heart of why separation holds for communication with a fidelity criterion.

\subsection{Recapitulation, speculation and further development for Part I}

Optimality of source-channel separation for communication with a fidelity criterion over a general, compound channel was proved. It was assumed that random codes are permitted. The probability of excess distortion criterion was used as the fidelity criterion. The distortion function was additive.

The proof used a layered black-box view-point. A proof due to Amos Lapidoth, based on the original separation argument of \cite{Shannon} is provided for a compound DMC. Various implications of the layered black-box proof technique are discussed. It is discussed, how these implications do not come out of the Shannon-Lapidoth proof technique.

Theorem \ref{KKUniXKK} is generalized to the unicast multi-user set-up in Part II. The Shannon-Lapidoth proof technique is not operational and the layered black-box proof technique is semi-operational. A fully operational proof is provided in Part III. This proof in Part III also gets to the heart of the question, ``\emph{Why} is a channel which is capable of communicating a source to within a certain distortion level, capable of communicating bits at any rate less  than the infimum of the rates needed to code the source to within the distortion level'': this lies at the heart of optimality of separation for communication with a fidelity criterion.

\section{Part II: multi-user setting}

\subsection{Introduction to, and contribution of Part II}
In Part II, the optimality of source-medium separation for multi-user communication with  fidelity criteria over a general, compound medium in the unicast set-up is proved. This generalizes the result of Part I. Unicast set-up means that the sources which various users want to communicate to each other are independent of each other. This will be a simple generalization of Part I. As in Part I, random codes are permitted. As in Part I, universality of the result refers to the fact that the medium model is compound. Further, as is the case in Part I, universality is over the medium, not the source.

Separation in Part II refers to source-medium separation as opposed to the separation of channel-coding and network-coding found  in the network-coding literature \cite{Koetter}.

In \cite{Tian}, the optimality of separation for communication with a fidelity criterion over a finite memory medium is proved. The result in Part II is a generalization of \cite{Tian}:  it holds for a general medium \emph{and} compound medium with the difference that \cite{Tian} uses the expected distortion criterion whereas Part II uses the probability of excess distortion criterion. The theorem in Part II  was proved simultaneously and independently of \cite{Tian}. A generalization to the compound setting is needed because 
the action of real media like wireless medium or the internet cannot be modeled as a known transition probability and one way to model these media might be that they belong to a set of transition probabilities.

\cite{Gastpar} provides examples where source-medium separation is not optimal for communication with fidelity criteria when the sources are correlated. Thus, the unicast setting is the extent to which the optimality of separation for communication with fidelity criteria can be expected to hold in a general framework.

\subsection{Notation and definitions for Part II}
Notation and definitions from Subsection \ref{NotAndDef} of Part I will be used. Recall, in particular, the superscript notation.

There are $N$ users: users $i$, $1 \leq i \leq N$. The medium input space at user $i$ is $\mathcal I_i$ and the medium output space at user $i$ is $\mathcal O_i$. $\mathcal I_i, 1 \leq i \leq N$ and $\mathcal O, 1 \leq i \leq N$ are finite sets. The medium is a sequence $m = <m^n>_1^\infty$ where 
\begin{align}
m^n &:  \prod_{i = 1}^N \mathcal I_i^n  \rightarrow \mathcal P \left ( \prod_{i = 1}^N \mathcal O_i^n \right ) \nonumber \\
         &: (\iota_i^n, 1 \leq i \leq N) \rightarrow m^n(\cdot | (\iota_i^n, 1 \leq i \leq N) )
\end{align}
When the block-length is $n$, the medium acts as $m^n(\cdot  | \cdot)$; $m^n(o_i^n, 1 \leq i \leq N|\iota_i^n, 1 \leq i \leq N)$ is the probability that the medium output at user $i$ is  $o_i^n$, $1 \leq i \leq N$  if the medium input at user $i$ is $\iota_i^n$, $1 \leq i \leq N$. A compound medium  is a set $\mathcal B$ of media with input space $\mathcal I_i$ at user $i$, $1 \leq i \leq N$, and output space $\mathcal O_i$ at user $i$, $1 \leq i \leq N$, and is denoted by $m \in \mathcal B$.

The source input spaces are $\mathcal X_{ij}, 1 \leq i, j \leq N$ and source output spaces are $\mathcal Y_{ij}, 1 \leq i, j \leq N$, respectively. $\mathcal X_{ij}, \mathcal Y_{ij}$ are finite sets.$X_{ij} = <X^n_{ij}>_1^\infty$ is the i.i.d. $X_{ij}$ source: $X_{ij}^n$ is the i.i.d. $X_{ij}$ sequence of length $n$. The i.i.d. $X_{ij}$ source needs to be communicated from user $i$ to user $j$. $d_{ij}: \mathcal X_{ij} \times \mathcal Y_{ij} \rightarrow [0, \infty )$ is the distortion function for communication from user $i$ to user $j$. The $n$-letter distortion function is defined additively: 
\begin{align}
d_{ij}^n(x_{ij}^n, y_{ij}^n) \triangleq \sum_{t=1}^n d_{ij}(x_{ij}^n(t), y_{ij}^n(t))
\end{align}
Communication of the sources $X_{ij}$ over $m \in \mathcal B$ requires modems $h_i = <h_i^n>_1^\infty$ at user $i$. The modems are allowed to collectively generate random codes; such random codes can be realized if there is a shared continuous valued random variable independent of all other random variables in the system available at all the modems. The exact model of modems is irrelevant; it suffices to say that the interconnection of medium and modems can be thought of as a different compound medium $w = <w^n>_1^\infty \in \mathcal M_{\mathcal B}$:
\begin{align}
w^n:  \prod_{i,j = 1}^N \mathcal X_{ij}^n  \rightarrow \mathcal P \left ( \prod_{i, j = 1}^N \mathcal Y_{ij}^n \right )
\end{align}
The inputs at user $i$ to $w \in \mathcal M_{\mathcal B}$ are $X_{ij} = <X_{ij}^n>_1^\infty$, $1 \leq j \leq N$. With these inputs, $w  \in \mathcal M_{\mathcal B}$ produces outputs $Y_{ji} = <Y_{ji}>_1^\infty$, $1 \leq j \leq N$ at user $i$.  Note the order of $i$ and $j$ in the notation: the reproduction of source $X_{ij}$ which is input at user $i$ and destined for user $j$ is $Y_{ij}$ at user $j$. Medium $m \in \mathcal B$ is said to be capable of communicating i.i.d. $X_{ij}$ source from user $i$ to user $j$, $1 \leq i, j \leq N$ to within a distortion $D_{ij}$ if there exist modems $<h_i^n>_1^\infty$ independent of the particular $m \in \mathcal B$ and non-negative real sequences $<\omega_{ij, n}>_1^\infty$, $\displaystyle {\lim_{n \to \infty} \omega_{ij, n} = 0}$ such that
\begin{align}
\Pr \left ( \frac{1}{n}d^n_{ij}(X^n_{ij}, Y^n_{ij}) > D_{ij} \right ) \leq \omega_{n, ij} \  \forall i, j, \ \forall m \in \mathcal B
\end{align}
In a separation architecture, each modem $<h_i^n>_1^\infty$ consists of a source modem $<h_{s, i}^n>_1^\infty$ and a medium modem $<h_{c, i}^n>_1^\infty$. 

The type of $<h_{s, i}^n>_1^\infty$ needed in Part II acts independently on sources $X_{ij}, 1 \leq j \leq N$ at user $i$ and message reproductions at user $i$:  $<h_{s, i}^n>_1^\infty$ encodes sources $X_{ij}, 1 \leq j \leq N$ to within distortion levels $D_{ij}$ and source decodes the message reproductions into source reproductions $Y_{ji}, 1 \leq j \leq N$. Thus, $<h_{s, i}^n>_1^\infty$ can be thought of as $n$ independent source encoders and $n$ independent source decoders. The infimum of rates needed to code the source $X_{ij}$ to within distortion $D_{ij}$ when the distortion function is $d_{ij}$ under the probability of excess distortion criterion
\begin{align}
\lim_{n \to \infty} \Pr \left ( \frac{1}{n}d^n_{ij}(X^n_{ij}, Y^n_{ij}) > D_{ij} \right ) = 0
\end{align}
is denoted by $R^P_{X_{ij}}(D_{ij})$. 

The interconnection of medium modems and the medium is used for reliable communication: a precise discussion is omitted and will be clear when discussing the theorem and its proof.

The question is: if there exists some architecture to communicate the i.i.d. $X_{ij}$ sources to within distortions $D_{ij}$ over $m \in \mathcal B$, does there exist a separation architecture, too? Under certain assumptions, this question is answered in the affirmative in the next section.

\subsection{Optimality of separation for multi-user communication with fidelity criteria over a general, compound medium in the unicast setting}

This section generalizes Theorem \ref{KKUniXKK} of Part I. 
\begin{theorem}[Optimality of separation: unicast, multi-user setting]
\label{MUUSCSMultiUserSettingMU}
Assume that random codes are permitted. Let $m \in \mathcal B$ be  capable of universally communicating i.i.d. $X_{ij}$ source from user $i$ to user $j$, $1 \leq i, j \leq N$ to within distortion levels $D_{ij}$ under additive distortion functions $d_{ij}$. The sources $X_{ij}$ are independent of each other. Then, reliable communication can be accomplished from user $i$ to user $j$, $1 \leq i, j \leq N$ over $m \in \mathcal B$ at rates $R_{ij} < R^P_{X_{ij}}(D_{ij})$. This reliable communication can be accomplished by consumption of same or lesser medium resources at each user as the medium resource consumption in the original architecture for communication of i.i.d. $X_{ij}$ sources to within distortion levels $D_{ij}$.

Further, if reliable communication can be accomplished over  $m \in \mathcal B$ from user $i$ to user $j$, $1 \leq i, j \leq N$ at certain rates strictly larger than $R^P_{X_{ij}}(D_{ij})$, then the independent, i.i.d. $X_{ij}$ sources  can be communicated from user $i$ to user $j$, $1 \leq i, j \leq N$ to within distortion levels $D_{ij}$ over $m \in \mathcal B$ by use of a separation architecture. The medium resource consumption in this separation architecture at each user is the same as or lesser than the medium resource consumption in the architecture for reliable communication at rate strictly $>R^P_{X_{ij}}(D_{ij})$ from user $i$ to user $j$ when all messages for transmission between all pairs of users $(i, j)$ are independent of each other and the distribution on every message is uniform.
\end{theorem}

\begin{IEEEproof}
i.i.d $X_{ij}$ sources can be communicated from user $i$ to user $j$, $1 \leq i, j \leq N$ to within distortion $D_{ij}$ with the help of modems $<h_i^n>_1^\infty$ at user $i$, $1 \leq i \leq N$. Consider two particular users, user $s$ and user $r$ and the communication from user $s$ to user $r$, neglecting all other users. The communication of i.i.d. $X_{sr}$ source from user $s$ to user $r$ to within distortion $D_{sr}$ under additive distortion function $d_{sr}$ can be thought of as point-to-point communication. By Theorem \ref{KKUniXKK}, reliable communication can be accomplished over $m \in \mathcal B$ from user $s$ to user $r$ at rates  $< R^P_{X_{sr}}(D_{sr})$ by user of an encoder-decoder $<E_{sr}^n, F_{sr}^n>_1^\infty$ which layers on top of $w \in \mathcal M_{\mathcal B}$. $E_{sr}^n$ generates i.i.d. $X_{sr}$ codes. Thus, the distribution of the inputs to $w \in \mathcal M_{\mathcal B}$ in the new architecture is the same \emph{in distribution} as the old architecture. In particular, this implies that the communication of sources $X_{ij}$, $(i, j) \neq (s, r)$ is unaffected in distribution in the new architecture, and thus, further in particular, for $(i, j) \neq (s, r)$, $X_{ij}$ is communicated to within distortion $D_{ij}$ in the new architecture. The process can be repeated inductively for all pairs of users $(i, j)$ and thus, reliable communication can be accomplished from user $i$ to user $j$, $1 \leq i, j \leq N$ over $m \in \mathcal B$ at rates $R_{ij} < R^P_{X_{ij}}(D_{ij})$. The argument concerning medium resource consumption in the architecture for reliable communication is the same as the argument in the proof of Theorem \ref{KKUniXKK} and is omitted. The proof of the second part of the theorem concerning existence of a separation architecture and the resource consumption in the separation architecture is the same as in the proof of Theorem \ref{KKUniXKK} and is omitted.
\end{IEEEproof}

\subsection{Discussion: partial applicability to the traditional wireless telephony problem}

The traditional wireless telephony problem is the following: there are $2N$ users $s_i, 1 \leq i \leq N$ and $s'_i, 1 \leq i \leq N$. User $s_i$ wishes to talk to user $s'_i$, $1 \leq i \leq N$. The voice signal at user $s_i$ is $V_i$ and the voice signal at user $s'_i$ is $V'_i$. The question is: how should architectures be designed to maximize the number of pairs of users which can communicate at the same time under certain constraints on resource consumption. 

Note
\begin{itemize}
\item
Voice signals are pairwise independent: $V_i$ is independent of $V_j, V'_j$ for $j \neq i$. However, $V_i$ and $V'_i$ are dependent
\item
Wireless medium is time varying and only partially known
\item
Voice admits distortion
\item
Other concerns: For example, security and delay
\end{itemize}
Assume
\begin{itemize}
\item
Voice signals $V_i, V'_j, 1 \leq i, j \leq N$ are independent, not just pairwise independent. This assumption is clearly incorrect but is made
\item
Wireless medium can be modeled as discussed in Part II
\item
Distortion function for measuring the quality of voice transmission is additive. This assumption is clearly incorrect but is made here
\item
Voice can be modeled as a stationary ergodic process for which the result of Part II can be generalized
\item
Other concerns, for example, delay and security are neglected in this discussion
\end{itemize}
Then, it follows that assuming random codes are permitted, separation architectures are optimal for the traditional wireless telephony problem. 

As stated before, the assumption that voice signals $V_i, V'_j, 1 \leq i, j \leq N$ are independent does not hold. A simple problem to understand the question when signals are dependent is the following: there are two users, user $s$ and user $s'$. User $s$ wishes to communicate source $V$ to user $s'$ under an additive distortion function $d$ and user $s'$ wishes to communicate source $V'$ to user $s$ under an additive distortion function $d'$. Sources $V$ and $V'$ might be dependent. Then, does separation hold, and if not, does separation hold to some extent? As stated before, \cite{Gastpar} gives examples where optimality of separation does not hold if the sources are correlated. In the above example, the authors do not expect the optimality of separation to hold in general. A question to understand is whether approximate optimality of separation in the sense, for example, of \cite{Tian}, holds, in this example.

\subsection{Recapitulation for Part II}

Optimality of source-medium separation for multi-user communication with fidelity criteria over a general, compound medium in the unicast setting was proved, thus generalizing Part I. It was assumed that random codes are permitted. The probability of excess distortion criterion was used as the fidelity criterion. The proof is a simple generalization of Part I. Partial applicability to the traditional wireless telephony problem was discussed.

A question to investigate is whether random codes are needed or not if $\mathcal B$ is compact.

The proof technique of interconnecting sub-systems by maintaining marginals is reminiscent of the behavioral, interconnection view,  \cite{Willems, WillemsCSM}, although in a stochastic setting.

\section{Part III: Why does separation hold?}

\subsection{Introduction to, and contribution of Part III}

Optimality of separation for communication with a fidelity criterion was proved by Shannon in \cite{Shannon} and generalized to the compound setting in Part I. The non-trivial step in the proofs in \cite{Shannon} and Part I is to prove the statement,

``A channel which is capable of communicating the i.i.d. $X$ source to within a certain distortion level is also capable of communicating bits reliably at any rate less than the infimum of the rates needed to code the i.i.d. $X$ source to within the same distortion level.''

Neither the proof in \cite{Shannon}, nor the proof in Part I get to the heart of why this statement is true. They provide proofs but not a deep understanding: the reason for this is the crucial reliance of the proofs on mutual-information expressions for rate-distortion function or the capacity of a channel, as discussed in Part I. Part III gets to the heart of why this statement is true. The channel-coding problem for obtaining rates of reliable communication can be thought of as a packing problem and the source-coding problem of obtaining rates needed for source compression can be thought of as a covering problem. The perspective taken in Part III is that of a \emph{randomized covering-packing} perspective. In contrast to the discussion in Part I where the nature of the proof was semi-operational, the perspective in Part III is \emph{fully} operational.

The view in Part III  is \emph{not} to get to general conditions under which separation holds and under which separation does not hold: this view of getting to general conditions is taken, for example, in \cite{VembuVerduSteinberg}. The view in Part III is to make the relevant assumptions needed in order to get to a  conceptual, intuitive understanding of why separation holds. In order to get to this understanding, the uniform $X$ source will be used instead of the i.i.d. $X$ source in the arguments. Uniform $X$ source puts uniform distribution on sequences with type \emph{precisely} $p_X$, compared to the i.i.d. $X$ source which puts ``most'' of the mass on sequences with type ``close to'' $p_X$. The use of the uniform $X$ source, because of a single type class, avoids $\epsilon$-$\delta$  arguments, and helps get to the essence of the optimality of separation for communication with a fidelity criterion.

The setting in Part III is the same as the setting in Part I with the following differences: 
\begin{enumerate}
\item
The uniform $X$ source is used instead of the i.i.d. $X$ source
\item
The distortion function is assumed to be permutation invariant instead of additive. Permutation invariant distortion functions are defined in Section \ref{NotAndDefIII}; additive distortion functions are a special case of permutation invariant distortion functions
\item
A technical condition, stated in Theorem \ref{KKUniUniKK}, is assumed on the rate-distortion function
\end{enumerate}

\subsection{Notation and definitions for Part III} \label{NotAndDefIII}

Notation and definitions from Subsection \ref{NotAndDef} of Part I will be used. Recall, in particular, the superscript notation.

The source input space is $\mathcal X$ and the source reproduction space is $\mathcal Y$. $\mathcal X$ and $\mathcal Y$ are finite sets. $X$ is a random variable on $\mathcal X$. Let $p_X(x)$ be rational $\forall x$. Let $n_0$ denote the least positive integer for which $n_0p_X(x)$ is an integer $\forall x \in \mathcal X$. Let $\mathcal U^n$ denote the set of sequences with (\emph{exact}) type $p_X$. $\mathcal U^n$ is non-empty if and only if $n_0$ divides $n$. Let $n' \triangleq n_0n$. Let $U^{n'}$ denote a random variable which is uniform on $\mathcal U^{n'}$ and zero elsewhere. Then, $<U^{n'}>_1^\infty$ is the uniform $X$ source and is denoted by $U$. The uniform $X$ source can be defined only for those $X$ for which $p_X(x)$ is rational $\forall x \in \mathcal X$.

Let $q$ denote a type on the set $\mathcal Y$ which is achievable when the block-length is $n'$. $\mathcal V_q^{n'}$ is the set of all sequences with type $q$. The uniform distribution on $\mathcal V_q^{n'}$ is $V_q^{n'}$.

Since the uniform $X$ source is defined only for block-lengths $n'$, distortion function, channels, encoders and decoders will be defined only for block-lengths $n'$.

$d = <d^{n'}>_1^\infty$ is the distortion function where $d^{n'}: \mathcal X^{n'} \times \mathcal Y^{n'} \rightarrow [0, \infty )$. Let $\pi^{n'}$ be a permutation (rearrangement) of $(1, 2, \ldots, n')$. That is, for $1 \leq i \leq n'$, $\pi^{n'}(i) \in \{1, 2, \ldots, n' \}$ and that, $\pi^{n'}(i)$, $1 \leq i \leq n'$ are different. For $x^{n'} \in \mathcal X^{n'}$, denote 
\begin{align}
\pi^{n'}x^{n'} \triangleq (x^{n'}(\pi^{n'}(1)), x^{n'}(\pi^{n'}(2)), \ldots, x^{n'}(\pi^{n'}(n')))
\end{align}
For $y^{n'} \in \mathcal Y^{n'}$, $\pi^{n'}y^{n'}$ is defined analogously. $<d^{n'}>_1^\infty$ is said to be permutation invariant if $\forall n'$,
\begin{align}
d^{n'}(\pi^{n'}x^{n'}, \pi^{n'}y^{n'}) = d^{n'}(x^{n'}, y^{n'}), \forall x^{n'}\in \mathcal X^{n'}, y^{n'} \in \mathcal Y^{n'} 
\end{align}
The distortion function in Part III is assumed to be permutation invariant. An additive distortion function is an example of a permutation invariant distortion function.

The  channel is a sequence $k = <k^{n'}>_ 1^\infty$ where $k^{n'}$ is defined as in Part I. The compound channel is a set $\mathcal D$ of channels and is denoted by $k \in \mathcal D$. Encoder-decoder $<e^{n'}, f^{n'}>_1^\infty$ is defined as in Part I with the difference that definitions are made only for block-lengths $n'$. $k \in \mathcal D$ is said to communicate the uniform $X$ source to within a distortion $D$ if (\ref{PEDCriterionChannel}) is replaced with 
\begin{align} \label{PEDUniformCriterionChannel}
\Pr \left ( \frac{1}{n'}d^{n'}(U^{n'}, Y^{n'}) > D \right ) \leq \omega_{n'} \  \forall k \in \mathcal D
\end{align}
for some $<\omega_{n'}>_1^\infty$, $\omega_{n'} \to 0$ as $n' \to \infty$. 

Source-code $<e_s^{n'}, f_s^{n'}>_1^\infty$ is defined as in Part I with the difference that definitions are made only for block-lengths $n'$. $<e_s^{n'}, f_s^{n'}>_1^\infty$ is said to code the uniform $X$ source to within a distortion $D$ if (\ref{SourceCodeD}) is replaced with
\begin{align} \label{SourceCodeUniformD}
\lim_{n' \to \infty} \Pr \left ( \frac{1}{n'} d^{n'}(U^{n'}, Y^{n'}) > D \right ) = 0
\end{align}
The infimum of rates needed to code the uniform $X$ source to within a distortion $D$ is $R^P_U(D)$. A definition not made in Part I is that of the rate-distortion function under the $\inf$ probability of excess distortion criterion. This definition was not needed in Part I but is needed in Part III. $<e_s^{n'}, f_s^{n'}>_1^\infty$ is said to code the uniform $X$ source to within a distortion $D$ under the $\inf$ probability of excess distortion criterion if (\ref{SourceCodeUniformD}) is replaced with
\begin{align} \label{SourceCodeUniformInfD}
\liminf_{n' \to \infty} \Pr \left ( \frac{1}{n'} d^{n'}(U^{n'}, Y^{n'}) > D \right ) = 0
\end{align}
The infimum of rates needed to code the uniform $X$ source to within a distortion $D$ under the $\inf$ probability of excess distortion criterion is denoted by $R^P_U(D, \inf)$.

Channel-code $<e_c^{n'}, f_c^{n'}>_1^\infty$ is defined as in Part I with the difference that definitions are made only for block-lengths $n'$. Reliable achievability of rate $R$ over $k \in \mathcal D$ is defined as in Part I with the difference that only block-lengths $n'$ matter. Capacity of $k \in \mathcal D$ is the supremum of all reliably achievable rates. 

As in Part I, $<e_s^{n'} \circ e_c^{n'}, f_s^{n'} \circ f_c^{n'}>_1^\infty$ is the separation-based encoder-decoder.

As in Part I, the question is: if there exists some architecture to communicate the uniform $X$ source to within a distortion $D$ over $k \in \mathcal D$, does there exist a separation architecture too? This question is answered in the affirmative under certain assumptions in the next subsection; note, in particular, the technical assumption on $R^P_U(D)$ in the statement of Theorem \ref{KKUniUniKK}. The important point for Part III is that the proof gets to the essence of why separation holds for communication with a fidelity criterion.

\subsection{Optimality of separation for communication with a fidelity criterion over a general, compound channel}

\begin{theorem}[Optimality of separation] \label{KKUniUniKK}
Assume that random codes are permitted. Assume that $R^P_U(D, \inf) = R^P_U(D)$. Let $k \in \mathcal D$ be capable of communicating the uniform $X$ source to within a distortion level $D$ under a permutation invariant distortion function $d$. Then, reliable communication can be accomplished over $k$ at rates $<R^P_U(D)$. This reliable communication can be accomplished with consumption of channel resources same as the channel resource consumption in the original architecture which communicates the uniform $X$ source to within a distortion $D$ over $\mathcal D$.

Further, if reliable communication can be accomplished over $k \in \mathcal D$ at a certain rate strictly $>R^P_U(D)$, then the uniform $X$ source can be communicated to within a distortion $D$ over $k \in \mathcal A$ by use of a separation architecture. The channel resource consumption in this separation architecture is the same as the channel resource consumption in the architecture for reliable communication at rate strictly $>R^P_U(D)$ when the distribution on the message set is uniform.
\end{theorem}

\begin{IEEEproof}
$k \in \mathcal D$ is capable of communicating the uniform $X$ source to within a distortion $D$ with the help of some encoder-decoder $<e^{n'}, f^{n'}>_1^\infty$. Consider the channel set
\begin{align}
\mathcal C_{\mathcal D} \triangleq <e^{n'} \circ k^{n'} \circ f^{n'}>_1^\infty, k \in \mathcal D
\end{align}
$c = <c^{n'}>_1^\infty$ is a compound channel with input space $\mathcal X$ and output space $\mathcal Y$. It will be proved that by use of some encoder-decoder $<E^{n'}, F^{n'}>_1^\infty$, reliable communication can be accomplished over $c \in \mathcal C_{\mathcal D}$ at rates $<R^P_U(D, \inf)$ with consumption of same channel resources as the architecture $<e^{n'} \circ k^{n'} \circ f^{n'}>_1^\infty$, when used for communicating the uniform $X$ source to within a distortion $D$. By use of the assumption $R^P_U(D) = R^P_U(D, \inf)$, it will follow that by use of encoder-decoder $<E^n \circ e^n, f^n \circ F^n>_1^\infty$, reliable communication can be accomplished over $k \in \mathcal D$ at rates $<R^P_U(D)$ with consumption of same channel resources as the architecture $<e^{n'} \circ k^{n'} \circ f^{n'}>_1^\infty$, when used for communicating the uniform $X$ source to within a distortion $D$.

This will be done by use of parallel random-coding arguments for two problems:
\begin{itemize}
\item
\emph{Channel-coding problem:} 
Rates of reliable communication over $k \in \mathcal C_{\mathcal D}$.
\item
\emph{Source-coding problem:} 
Rates of coding  for the uniform $X$ source to within a distortion $D$ under the $\inf$ probability of excess distortion criterion.
\end{itemize}

\emph{Codebook generation:}
\begin{itemize}
\item
\emph{Codebook generation for the channel-coding problem:}
Let reliable communication be desired at rate $R$. Generate $2^{\lfloor n'R \rfloor}$ sequences independently and uniformly from $\mathcal U^{n'}$. This is the codebook $\mathcal K^{n'}$.
\item
\emph{Codebook generation for the source-coding problem:}
Let source-coding be desired at rate $R$. Generate $2^{\lfloor n'R \rfloor}$ codewords  independently and uniformly from $\mathcal V_q^{n'}$ for some type $q$ on $\mathcal Y$ which is achievable for block-length $n'$. This is the codebook $\mathcal L^{n'}$. 
\end{itemize}

\emph{Joint typicality:}

Joint typicality for both the channel-coding and source-coding problems is defined as follows:
$(u^{n'}, y^{n'}) \in \mathcal U^{n'} \times \mathcal Y^{n'}$ jointly typical if 
\begin{align}
\frac{1}{n'} d^{n'} (u^{n'}, y^{n'}) \leq D
\end{align}

\emph{Decoding and encoding:}
\begin{itemize}
\item
\emph{Decoding for the channel-coding problem:}
Let $y^{n'}$ be received. If there exists unique $u^{n'} \in \mathcal K^{n'}$ for which $(u^{n'}, y^{n'})$ jointly typical, declare that $u^{n'}$ is transmitted, else declare error. 
\item
\emph{Encoding for the source-coding problem:}
Let $u^{n'} \in \mathcal U^{n'}$ need to be source-coded. If there exists some $y^{n'} \in \mathcal L^{n'}$, encode $u^{n'}$ to one such $y^{n'}$, else declare error.
\end{itemize}

\emph{Some notation:}
\begin{itemize}
\item
\emph{Notation for the channel-coding problem:} 
Let message $m^{n'} \in \mathcal M_R^{n'}$ be transmitted. Codeword corresponding to $m^{n'}$ is $u_c^{n'}$. Non-transmitted  codewords are ${u'}_1^{n'}, {u'}_2^{n'}, \ldots, {u'}_{2^{\lfloor n'R \rfloor} - 1}^{n'}$. $u_c^{n'}$ is a realization of $U_c^{n'}$. $U_c^{n'}$  is uniform on $\mathcal U^{n'}$. ${u'}_i^{n'}$ is a realization of ${U'}_i^{n'}$. ${U'}_i^{n'}$ is uniform on $\mathcal U^{n'}$, $1 \leq i \leq 2^{\lfloor n'R \rfloor} - 1$. $U_c^{n'}, {U'_i}^{n'}, 1 \leq i \leq 2^{\lfloor n'R \rfloor} - 1$ are independent of each other. The channel output is $y^{n'}$. $y^{n'}$ is a realization of $Y^{n'}$. $y^{n'}$ may depend on $u_c^{n'}$ but does not depend on ${u'_i}^{n'}, 1 \leq i \leq 2^{\lfloor n'R \rfloor} - 1$.  As random variables, $Y^{n'}$ and $U_c^{n'}$ might be dependent but $Y^{n'}, {U'_i}^{n'}, 1 \leq i \leq 2^{\lfloor n'R \rfloor} - 1$ are independent. If the type $q$ of the sequence $y^{n'}$ needs to be explicitly denoted, the sequence is denoted by $y_q^{n'}$. $\mathcal G^{n'}$ is the set of all achievable types $q$ on $\mathcal Y$ for block-length $n'$. $Y^{n'}$ may depend on the channel $c \in \mathcal C_{\mathcal A}$; this dependence is suppressed. 

\item
\emph{Notation for the source-coding problem:}
$u_s^{n'}$ is the sequence which needs to be source-coded. $u_s^{n'}$ is a realization of $U_s^{n'}$ which is uniformly distributed on $\mathcal U^{n'}$. The codewords are $y_{q,i}^{n'}, 1 \leq i \leq 2^{\lfloor n'R \rfloor}$ where $q$ denotes the type. $y_{q, i}^{n'}$ is a realization of $V_{q, i}^{n'}, 1 \leq i \leq 2^{\lfloor n'R \rfloor}$ where $V_{q, i}^{n'}$ is uniformly distributed on the subset of $\mathcal Y^{n'}$ consisting of all sequences with type $q$. $u_s^{n'}, y_{q, i}^{n'}, 1 \leq i \leq 2^{\lfloor n'R \rfloor}$ are independently generated; as random variables, $U_s^{n'}, Y_{q, i}^{n'}, 1 \leq i \leq 2^{\lfloor n'R \rfloor}$ are independent. $\mathcal G^{n'}$ is the set of all achievable types $q$ on $\mathcal Y$ for block-length $n'$
\end{itemize}

\emph{Error analysis:}
For the channel-coding problem, the probability of correct decoding is analyzed and for the source-coding problem, the probability of error is analyzed.
\begin{itemize}
\item
\emph{Error analysis for the channel-coding problem:}
From the encoding-decoding rule, it follows that the event of correct decoding given that a particular message is transmitted is 
\begin{align}\label{CorrectDecodingEvent}
& \left \{ \frac{1}{n'} d^{n'}(U_c^{n'}, Y^{n'}) \leq D \right \} \cap \nonumber \\
& \hspace{2cm} \cap_{i = 1}^{2^{\lfloor n'R \rfloor} - 1} \left \{ \frac{1}{n'} d^{n'} ({U'}_i^{n'}, Y^{n'}) > D  \right \}
\end{align}
\item
\emph{Error analysis for the source-coding problem:}
From the encoding-decoding rule, it follows that the error event given that a particular message needs to be source-coded is
\begin{align}\label{ErrorEvent}
\cap_{i=1}^{2^{\lfloor n'R \rfloor}} \left \{  \frac{1}{n'}d^{n'}(u^{n'}, V_{q, i}^{n'}) > D \right \}
\end{align}
\end{itemize}
Note that there is choice of $q$ for codebook generation.

\emph{Calculation:}
\begin{itemize}
\item
\emph{Calculation of the probability of correct decoding for the channel-coding problem:}

A bound on the probability of the correct decoding event (\ref{CorrectDecodingEvent}) is calculated, using essentially standard arguments for calculating such bounds, in Appendix \ref{CalculationAppendix} and is 
\begin{align}
\geq &  -\omega_{n'} +  \nonumber \\ 
         & \hspace{1cm} \left [ \inf_{q \in \mathcal G^{n'}} \Pr \left ( \left \{ \frac{1}{n'} d^{n'} ({U}^{n'}, y_q^{n'})  \right . \right . \right . \nonumber \\ 
       & \hspace{2.9cm} \left . \textcolor{white}{\inf_{q \in \mathcal G^{n'}}} \left .  \left .  \textcolor{white}{\frac{1}{n'}}  > D \right \} \right ) \right ] ^{ 2^{\lfloor n'R \rfloor } - 1} \nonumber \\      
\end{align}
where $U^{n'}$ is uniform on $\mathcal U^{n'}$. 

Rate $R$ is achievable if 
\begin{align}
 &  -\omega_{n'} +  \nonumber \\ 
         & \hspace{1cm} \left [ \inf_{q \in \mathcal G^{n'}} \Pr \left ( \left \{ \frac{1}{n'} d^{n'} ({U}^{n'}, y_q^{n'})  \right . \right . \right . \nonumber \\ 
       & \hspace{2.9cm} \left . \textcolor{white}{\inf_{q \in \mathcal G^{n'}}} \left .  \left .  \textcolor{white}{\frac{1}{n'}}  > D \right \} \right ) \right ] ^{ 2^{\lfloor n'R \rfloor } - 1} \nonumber \\                
       & \hspace{5.2cm}          \to 1 \ \mbox{as} \ n' \to \infty
\end{align}
$\omega_{n'} \to 0$ as $n' \to \infty$. It follows that rate $R$ is achievable if 
\begin{align}\label{JCorrectCalculationJ}
          & \left [ \inf_{q \in \mathcal G^{n'}} \Pr \left ( \left \{ \frac{1}{n'} d^{n'} ({U}^{n'}, y_q^{n'}) > D \right \} \right ) \right ] ^{ 2^{\lfloor n'R \rfloor } - 1} \nonumber \\ 
          &  \hspace{5.2cm} \to 1 \ \mbox{as} \ n' \to \infty
\end{align}
\item
\emph{Calculation of probability of error for the source-coding problem:}

An expression for the probability of the error event (\ref{ErrorEvent}) is calculated, using standard arguments for calculating these probabilities, in Appendix \ref{CalculationAppendix} and is
\begin{align}
 \left [  \inf_{q \in \mathcal G^{n'}} \Pr \left (  \left \{  \frac{1}{n'}d^{n'}(u^{n'}, V_q^{n'}) > D \right \} \right )\right ] ^{2^{\lfloor n'R \rfloor }}
\end{align}
where $V_q^{n'}$ is uniform on $\mathcal V_q^{n'}$. This expression is calculated using essentially standar
The infimum in the above expression reflects the existing choice of type in the codeword generation process.

Since the $\inf$ probability of excess distortion criterion is used, it follows that rate $R$ is achievable if 
\begin{align}\label{JSourceErrorJ}
 & \left [  \inf_{q \in \mathcal G^{n'_i}} \Pr \left (  \left \{  \frac{1}{n'_i}d^{n'_i}(u^{n'_i}, V_q^{n'_i}) > D \right \} \right )\right ] ^{2^{\lfloor n'_iR \rfloor }}  \nonumber \\
 & \hspace{1cm}   \to 0 \ \mbox{for some} \ n'_i = n_0 n_i, \ n_i \  \to \infty
\end{align}
\end{itemize}

\emph{Connection between channel-coding and source-coding:}

The calculation required in the channel-coding problem is 
\begin{align} \label{CCCalc}
\inf_{q \in \mathcal G^{n'}}\Pr \left ( \left \{ \frac{1}{n'} d^{n'} ({U}^{n'}, y_q^{n'}) > D \right \} \right )
\end{align}
and the calculation required in the source-coding problem is
\begin{align} \label{SCCalc}
\inf_{q \in \mathcal G^{n'}} \Pr \left ( \left \{ \frac{1}{n'} d^{n'} ({u}^{n'}, V_q^{n'}) > D \right \} \right )
\end{align}
It will be proved that (\ref{CCCalc}) and (\ref{SCCalc}) are equal. It will be proved more generally that 
\begin{align}\label{JJMainDualityJJ}
& \Pr \left ( \left \{ \frac{1}{n'} d^{n'} ({U}^{n'}, y_q^{n'}) > D \right \} \right ) = \nonumber \\ 
& \hspace{3cm} \Pr \left ( \left \{ \frac{1}{n'} d^{n'} ({u}^{n'}, V_q^{n'}) > D \right \} \right )
\end{align} 
This is a symmetry argument and requires the assumption of permutation invariant distortion function. The idea is that the left hand side of (\ref{JJMainDualityJJ}) depends only on the type of $y_q^{n'}$. From this it follows that the left hand side of (\ref{JJMainDualityJJ}) is equal to 
\begin{align} \label{SymmetryExpression}
\Pr \left ( \left \{ \frac{1}{n'} d^{n'} ({U}^{n'}, V_q^{n'}) > D \right \} \right )
\end{align}
where $V_q^{n'}$ is independent of $U^{n'}$. Similarly, the right hand side of (\ref{JJMainDualityJJ}) depends only on the type of $u^{n'}$ and from this it follows that the right hand side of (\ref{JJMainDualityJJ}) is also equal to (\ref{SymmetryExpression}). (\ref{JJMainDualityJJ}) follows. Details are in Appendix \ref{CalculationAppendix}.

\emph{Proof that a channel which is capable of communicating the uniform $X$ source to within a certain distortion level is also capable of communicating bits reliably at any rate less than the infimum of the rates needed to code the uniform $X$ source to within the same distortion level under the $\inf$ probability of excess distortion criterion:}

Denote 
\begin{align}
& A_{n'} \triangleq \inf_{q \in \mathcal G^{n'}}\Pr \left ( \left \{ \frac{1}{n'} d^{n'} ({U}^{n'}, y_q^{n'}) > D \right \} \right ) = \nonumber \\ 
& \hspace{2cm}                                            \inf_{q \in \mathcal G^{n'}} \Pr \left ( \left \{ \frac{1}{n'} d^{n'} ({u}^{n'}, V_q^{n'}) > D \right \} \right )
\end{align}
From (\ref{JCorrectCalculationJ}), it follows that rate $R$ is achievable for the channel-coding problem if 
\begin{align} \label{ChannelCalculationCriterion}
 (A_{n'}) ^{ 2^{\lfloor n' R \rfloor } - 1} \to 1 \ \mbox{as} \ n' \to \infty
\end{align}
From (\ref{JSourceErrorJ}), it follows that rate $R$ is achievable for the source-coding problem if 
\begin{align}
& (A_{n'_i})^{2^{\lfloor n'_i R \rfloor }} \to 0 \ \mbox{as} \ n'_i \to \infty  \nonumber \\ 
& \hspace{1cm} \mbox{for some} \ n'_i = n_0 n_i\ \mbox{for some}\  n_i \to \infty
\end{align}
Let
\begin{align} \label{JDefAlphaJ}
\alpha \triangleq \sup \{ R \ | \ (\ref{ChannelCalculationCriterion}) \ \mbox{holds} \}
\end{align}
Then, if $R' > \alpha$, 
\begin{align}\label{FinalStretch1}
& \lim_{n'_i \to \infty} (A_{n'_i})^{2^{\lfloor n_i R' \rfloor} - 1} < 1\ \forall \  R' > \alpha  \nonumber \\ 
& \hspace{3cm}   \mbox{for some sequence} \ n'_i \to \infty
\end{align}
$n'_i$ may depend on $R'$.

Then,  
\begin{align}\label{FinalStretch2}
\lim_{n'_i \to \infty} (A_{n'_i})^{2^{\lfloor n'_iR'' \rfloor} - 1}  = 0 \ \mbox{for}\ R'' > R'
\end{align}
(\ref{FinalStretch1}) and (\ref{FinalStretch2}) hold for all $R'' > R' > \alpha$. It follows that rates larger than $\alpha$ are achievable for the source-coding problem. 

Thus, a channel which is capable of communicating the uniform $X$ source to within a certain distortion level is also capable of communicating bits reliably at any rate less than the infimum of the rates needed to code the uniform $X$ source to within the same distortion level under the $\inf$ probability of excess distortion criterion.

\emph{Wrapping up the proof of the theorem:}

It follows that if $k \in \mathcal D$ is capable of communicating the uniform $X$ source to within a distortion level $D$ under a permutation invariant distortion function $d$, then reliable communication can be accomplished over $k$ at rates $<R^P_U(D, \inf)$. By use of the assumption $R^P_U(D) = R^P_U(D, \inf)$, it follows that reliable communication can be accomplished over $k$ at rates $<R^P_U(D)$. The argument concerning channel resource consumption is the same as in the proof of Theorem \ref{KKUniXKK}. The argument for the second part of the theorem concerning existence of a separation architecture and channel resource consumption in the separation architecture is the same as in the proof of Theorem \ref{KKUniXKK}.

\end{IEEEproof}

\subsection{Discussion}

The proof of Theorem \ref{KKUniUniKK} can be viewed as exhibiting a randomized covering-packing perspective on the optimality of separation for communication with a fidelity criterion: the source-coding problem can be thought of as a covering problem and the channel-coding problem can be thought of as a packing problem; the proof uses a random-coding argument for each of these problems and draws a parallel between them. The essence of why separation holds is captured in (\ref{JJMainDualityJJ}). The reader should compare this proof with the Shannon-Lapidoth proof described in Part I.
The proof can also be viewed as a connection (duality) between source-coding and channel-coding. The proof only uses the operational meanings of reliable communication and source coding and not explicit functional simplifications, for example, mutual-information expressions, for channel capacity or the rate-distortion function; the extent to which functional simplifications are required are in the calculations needed to arrive at (\ref{CCCalc}) and (\ref{SCCalc}). A precise definition of an operational proof is not possible; however, it can be intuitively interpreted from the context in which the word is used.

The technical condition $R^P_U(D) = R^P_U(D, \inf)$ is made on the rate-distortion function.  It is unlikely that this technical condition will hold for an arbitrary permutation-invariant distortion function $d = <d^{n'}>_1^\infty$; however, the authors conjecture that this technical condition will hold for ``many well behaved'' permutation invariant distortion functions. The validity of this technical condition is proved \emph{operationally} for an additive distortion function in Chapter 5 of \cite{MukulPhdThesis}; recall that an additive distortion function is permutation invariant. 

The proof technique of Part III is generalized to the i.i.d. $X$ source for an additive distortion function in Chapter 5 of \cite{MukulPhdThesis}. The authors conjecture that the proof technique of Part III can be used for the i.i.d. $X$ source for many ``well behaved'' permutation invariant distortion functions. 

An alternate proof of the rate-distortion theorem which the authors believe is more fundamental than the original proof of Shannon is presented in Chapter 5 of \cite{MukulPhdThesis}.

\subsection{Recapitulation for Part III}
A perspective which gets to the heart of why separation holds for the problem of communication with a fidelity criterion is provided. The perspective is an operational, randomized covering-packing perspective.

\section{Recapitulation}

The abstract to this paper and the recapitulations to each of Part I, Part II, and Part III, together recapitulate this paper. The development in this paper is brief; an elaborate development of a large part of this paper and further discussions can be found in \cite{MukulPhdThesis}.

\bibliographystyle{IEEEtran}
\bibliography{togetherpaperBibliography.bib}  

\begin{thebibliography}{10}
\providecommand{\url}[1]{#1}
\csname url@samestyle\endcsname
\providecommand{\newblock}{\relax}
\providecommand{\bibinfo}[2]{#2}
\providecommand{\BIBentrySTDinterwordspacing}{\spaceskip=0pt\relax}
\providecommand{\BIBentryALTinterwordstretchfactor}{4}
\providecommand{\BIBentryALTinterwordspacing}{\spaceskip=\fontdimen2\font plus
\BIBentryALTinterwordstretchfactor\fontdimen3\font minus
  \fontdimen4\font\relax}
\providecommand{\BIBforeignlanguage}[2]{{%
\expandafter\ifx\csname l@#1\endcsname\relax
\typeout{** WARNING: IEEEtran.bst: No hyphenation pattern has been}%
\typeout{** loaded for the language `#1'. Using the pattern for}%
\typeout{** the default language instead.}%
\else
\language=\csname l@#1\endcsname
\fi
#2}}
\providecommand{\BIBdecl}{\relax}
\BIBdecl

\bibitem{CsiszarKorner}
I.~Csisz\'ar and J.~Korner, \emph{Information theory: coding theorems for
  discrete memoryless systems}.\hskip 1em plus 0.5em minus 0.4em\relax
  Akad\'emiai Kiad\'o, 1997.

\bibitem{VerduHan}
S.~Verdu and T.~S. Han, ``A general formula for channel capacity,'' \emph{IEEE
  Transactions on Information Theory}, vol. 40, issue 4, pp. \ pages
  1147--1157, July 1994.

\bibitem{Shannon}
C.~E. Shannon, ``Coding theorems for a discrete source with a fidelity
  criterion,'' \emph{Institute of Radio Engineers, National Convention Record},
  vol. 7, part 4, pp. 142--163, March 1959.

\bibitem{GallagerInformationTheory}
R.~G. Gallager, \emph{Information theory and reliable communication}.\hskip 1em
  plus 0.5em minus 0.4em\relax Wiley, January 1968.

\bibitem{ShannonReliable}
C.~E. Shannon, ``A mathematical theory of communication,'' \emph{Bell System
  Technical Journal}, vol.~27, pp. 379--423 (Part 1) and pp. 623--656 (Part 2),
  July (Part 1) and October (Part 2) 1948.

\bibitem{LomnitzFeder}
Y.~Lomnitz and M.~Feder, ``Communication over individual channels,'' \emph{IEEE
  transactions on information theory}, vol. 57, issue 11, pp. 7333 -- 7358,
  November 2011.

\bibitem{GallagerDigital}
R.~G. Gallager, \emph{Principles of digital communications}.\hskip 1em plus
  0.5em minus 0.4em\relax Cambridge University Press, 2008.

\bibitem{AgarwalSahaiMitterAllerton}
M.~Agarwal, A.~Sahai, and S.~K. Mitter, ``Coding into a source: A direct
  inverse rate-distortion theorem,'' in \emph{Proceedings of the 44th Annual
  Allerton Conference on Communication, Control, and Computing, 2006}, pp.
  569--578.

\bibitem{GrayPursley}
M.~B. Pursley and R.~M. Gray, ``Source coding theorems for stationary,
  continuous-time stochastic processes,'' \emph{Annals of Probability}, vol.~5,
  no.~6, pp. 966--986, 1977.

\bibitem{Ziv}
J.~Ziv, ``Coding of sources with unknown statistics - {I}{I}: Distortion
  relative to a fidelity criterion,'' \emph{IEEE Transactions on Information
  Theory}, vol. 18, issue 4, pp. 389--394, May 1972.

\bibitem{Ziv2}
------, ``Distortion-rate theory for individual sequences,'' \emph{IEEE
  Transactions on Information Theory}, vol. 26, issue 2, pp. 137--143, March
  1980.

\bibitem{MukulSwastikMitter}
M.~Agarwal, S.~Kopparty, and S.~K. Mitter, ``The universal capacity of channels
  with given rate-distortion in the absence of common randomness,'' in
  \emph{Proceedings of the 47th Annual Allerton Conference on Communication,
  Control, and Computing, 2009}, pp. 700--707.

\bibitem{Koetter}
R.~Koetter, M.~Effros, and M.~Medard, ``On a theory of network equivalence,''
  in \emph{Proceedings of the IEEE Information Theory Workshop on Networking
  and Information Theory, 2009}, pp. 326 -- 330.

\bibitem{Tian}
C.~Tian, J.~Chen, S.~Diggavi, and S.~Shamai, ``Optimality and approximate
  optimality of source-channel separation in networks,'' \emph{Submitted to
  IEEE Transactions on Information Theory, arXiv:1004.2648}.

\bibitem{Gastpar}
M.~Gastpar, ``To code or not to code,'' Ph.D. dissertation, \'Ecole
  Polytechnique F\'ed\'erale de Lausanne, 2002.

\bibitem{Willems}
J.~C. Willems, ``Models for dynamics,'' \emph{Dynamics Reported}, vol.~2, pp.
  171--269, April 1989.

\bibitem{WillemsCSM}
------, ``The behavioral approach to open and interconnected systems,''
  \emph{IEEE Control Systems Magazine}, vol.~27, pp. 46--99, December 2007.

\bibitem{VembuVerduSteinberg}
S.~Vembu, S.~Verdu, and Y.~Steinberg, ``The source-channel separation theorem
  revisited,'' \emph{IEEE Transactions on Information Theory}, vol. 41, issue
  1, pp. 44--54, January 1995.

\bibitem{MukulPhdThesis}
M.~Agarwal, ``A universal, operational theory of multi-user communication with
  fidelity criteria,'' Ph.D. dissertation, Massachusetts Institute of
  Technology, February 2012.

\bibitem{CoverThomas}
T.~M. Cover and J.~A. Thomas, \emph{Elements of Information Theory},
  2nd~ed.\hskip 1em plus 0.5em minus 0.4em\relax Wiley-Interscience, 2006.

\end{thebibliography}

\appendices 

\section{Error analysis in the proof of Theorem \ref{KKUniXKK}} \label{AppendixErrorAnalysis}
Denote: $<g^n>_1^\infty = <E^n \circ c^n \circ F^n>_1^\infty$.

Denote: 
\begin{itemize}
\item $x^n \in \mathcal K^n$ is the transmitted codeword corresponding to a a particular message $m^n$
\item $y^n \in \mathcal Y^n$ is the received sequence 
\item $z^n \in \mathcal K^n$ is a codeword which is \emph{not} transmitted
\end{itemize}
$g^n(\{m^n\}^c|m^n)$ needs to be calculated.

\begin{align}
g^n(\{m^n\}^c|m^n) \subset  \Pr(\mathcal E_1^n  \cup \mathcal E_2^n)
                                                                     \leq \Pr(\mathcal E_1^n) + \Pr(\mathcal E_2^n)
\end{align}
where 
\begin{itemize}
\item
$\mathcal E_1^n$:  $(x^n, y^n)$ is not $\epsilon$ jointly typical
\item
$\mathcal E_2^n$: $\exists z^n$ such that $(z^n, y^n)$ is $\epsilon$ jointly typical
\end{itemize}

By definition of $c \in \mathcal C_{\mathcal A}$, $\Pr(\mathcal E_1^n) \to 0$ as $n \to \infty$ at a uniform rate over $c \in \mathcal C_{\mathcal A}$, and independently of $m^n$.

Calculation of $\Pr(\mathcal E_2^n)$ requires a method of types calculation, see \cite{CsiszarKorner} or \cite{CoverThomas}, similar in arguments to those used in the proof of Lemma 1 in \cite{LomnitzFeder} which, in turn, are similar to the error analysis in the conference version \cite{AgarwalSahaiMitterAllerton} of Part I. This calculation is carried out below:

Fix $y^n$. Let $y^n$ have type $q_Y$.  Let $z^n$ be a particular non-transmitted codeword. The distribution $Z^n$ on $z^n$ is generated i.i.d. $X$. $Z^n$ and $y^n$ are independent of each other. Let $q_{Z|Y}$ be a conditional distribution $\mathcal Y \rightarrow \mathcal P(\mathcal X)$. $q_{ZY}$ is the joint distribution resulting from $q_Y$ and $q_{Z|Y}$ Then,
\begin{align}
        & \Pr \left ( p_{Z^n|y^n} = q_{Z|Y} | p_{y^n} = q_Y \right ) \nonumber \\
\leq & \prod_{y \in \mathcal{Y}} 2^{-nq_Y(y)D(q_{Z|Y}(\cdot|y)||p_X)} \nonumber \\
=     & 2^{-n \sum_{y \in \mathcal{Y}}q_Y(y)D(q_{Z|Y}(\cdot|y||p_X)} \nonumber \\
=     & 2^{-n D(q_{ZY}||p_X q_Y)}
\end{align}
It follows by
\begin{itemize}
\item
noting that $q_Y$ is arbitrary,
\item
number of joint types on $\mathcal X \times \mathcal Y$ is $\leq (n+1)^{|\mathcal X||\mathcal Y|}$,
\item
and by use of the union bound,
\end{itemize}
that 
\begin{align}\label{FinalE2Bound}
  \Pr(\mathcal E_2^n) \leq (n+1)^{|\mathcal{X}||\mathcal{Y}|} 2^{\lfloor nR \rfloor}  2^{-n \inf _ {q_{ZY}\in \mathcal{T}} D(q_{ZY}||p_X q_Y) }
\end{align}
where
\begin{align}
\mathcal{T} \triangleq  \left \{ q_{ZY}: 
       \begin{array}{l} q_Z \in p_X \pm \epsilon \\
         \sum_{x \in \mathcal X, y \in \mathcal Y} q_{ZY}(x, y)d(x, y) \leq D \\
       \end{array} \right \}
\end{align}
The above bound on $\Pr(\mathcal E_2^n)$ is independent of $m^n$ and $c \in \mathcal C_{\mathcal A}$. As stated before, $\Pr(\mathcal E_1^n) \to 0$ as $n \to \infty$ at a uniform rate over $c \in \mathcal C_{\mathcal A}$, and independently of $m^n$. 

Thus, rates $R < \inf_ {q_{ZY} \in \mathcal T} D(q_{ZY}||p_X q_Y)$ are reliably achievable over $c \in \mathcal C_{\mathcal A}$. Now,
\begin{align}
& D(q_{ZY}||p_X q_Y) =  \nonumber \\ 
& \hspace{1cm} D(q_{Z}||p_X) + D(q_{ZY}||q_Z q_Y)  \geq    D(q_{ZY}||q_Z q_Y)
\end{align}
Thus, rates 
\begin{align}
R < \inf_ {q_{ZY} \in \mathcal T} D(q_{ZY}||q_Z q_Y) =  \inf_ {q_{ZY} \in \mathcal T} I(Z;Y)
\end{align}
are reliably achievable over $c \in \mathcal C_{\mathcal A}$. Now,
\begin{align}
 \inf_{q_{ZY} \in \mathcal T} I(Z;Y) =  \inf_{Z \in \mathcal T(p_X, \epsilon)} R^I_Z(D)
\end{align}
Thus, rates 
\begin{align}
R < \inf_{Z \in \mathcal T(p_X, \epsilon)} R^I_Z(D) 
\end{align}
are reliably achievable over $c \in \mathcal C_{\mathcal A}$. $\epsilon > 0$ is arbitrary and $R^I_X(\cdot)$ is continuous. Thus, rates $R <  R^I_X(D)$ are reliably achievable over $c \in \mathcal C_{\mathcal A}$. $R^I_X(D) = R^P_X(D)$. Thus, rates $<R^P_X(D)$ are reliably achievable over $c \in \mathcal C_{\mathcal A}$.

\section{Shannon-Lapidoth view of the optimality of separation for communication with a fidelity criterion over a compound DMC} \label{ShannonLapidothView}

Let $k = <k^n>_1^\infty \in \mathcal A$ be a compound DMC which is capable of communicating the i.i.d. $X$ source under the expected distortion criterion. As stated in Subsection \ref{LBBImplications}, it will be proved that $C^I(k \in \mathcal A) \geq R^I_X(D)$ from which it will follow that the capacity of $k \in \mathcal A$ is $\geq R^E_X(D)$. This will prove the equivalent of the first part of Theorem \ref{KKUniXKK} for a compound DMC. An argument will be made concerning channel resource consumption. The equivalent of second part follows by source-coding followed by channel-coding, as in the proof of Theorem \ref{KKUniXKK}. 

To prove $C^I(k \in \mathcal A) \geq R^I_X(D)$, it will be proved that $\forall n$, $C^I(k \in \mathcal A) \geq R^I_X(D + \omega_n)$. By the continuity of $R^I_X(\cdot)$, it will follow that $C^I(k \in \mathcal A) \geq R^I_X(D)$. To prove $C^I(k \in \mathcal A) \geq R^I_X(D)$, it will be proved that 
\begin{align} \label{AmosStep1SubStep1}
nR^I_X(D + \omega_n) \leq \inf_{k \in \mathcal A}I(X^n; Y^n)
\end{align} 
and 
\begin{align}\label{AmosStep1SubStep2}
\inf_{k \in \mathcal A}I(X^n; Y^n) \leq nC^I(k \in \mathcal A)
\end{align}
In (\ref{AmosStep1SubStep1}) and (\ref{AmosStep1SubStep2}), $Y^n$ may depend on the particular $k \in \mathcal A$; this dependence is suppressed in the notation. It is this dependence on $k \in \mathcal A$ that the infimum of the mutual informations is taken over.

If $\mathcal A$ is a singleton, (\ref{AmosStep1SubStep1}) is proved in \cite{Shannon}. If $\mathcal A$ is not a singleton, (\ref{AmosStep1SubStep1}) then follows by taking an infimum over all $k \in \mathcal A$.

To prove (\ref{AmosStep1SubStep2}), with input $X^n$ to the encoder $e^n$, denote the input to $k^n$ by $I^n$ and output of $k^n$ by $O^n$. $I^n$ is independent of $k \in \mathcal A$ and $O^n$ might depend on $k \in \mathcal A$. Fix a particular $k \in \mathcal A$. By the data processing inequality, 
\begin{align}
I(X^n; Y^n) \leq I(I^n; O^n)
\end{align}
Define, for $\iota \in \mathcal I$,
\begin{align} \label{TDistribution}
T(\iota) \triangleq \frac{1}{n} \sum_{t = 1 }^n \Pr(I^n(t) = \iota)
\end{align}
Since $e^n$ is independent of the particular $k \in \mathcal A$, $\Pr(I^n(t) = \iota)$ is independent of the particular $k \in \mathcal A$. Thus, $T(\cdot)$ is independent of the particular $k \in \mathcal A$.

Then,
\begin{align}
        & I(I^n; O^n) \nonumber \\
=      & H(O^n) - H(O^n|I^n) \nonumber \\
\leq  & \sum_{t = 1}^n H(O^n(t)) - H(O^n|I^n) \nonumber \\
& \hspace{1cm} \mbox{(by using $H(P, Q) = H(P) + H(Q|P)$ and}  \nonumber \\
& \hspace{4.8cm}\mbox{$H(P|Q, R) \leq H(P|Q)$) } \nonumber \\
=      & \sum_{t = 1}^n H(O^n(t)) - \nonumber \\
& \hspace{1.2cm}  \sum_{t = 1}^n H(O^n(t) | I^n, O^n(1), O^n(2), \ldots, O^n(t-1)) \nonumber \\
=      & \sum_{t = 1}^n H(O^n(t)) - \sum_{t = 1}^n H(O^n(t) | I^n(t)) \nonumber \\ 
& \ \ \ \ \ \ \ \ \ \ \ \ \ \ \ \ \ \ \ \ \ \ \ \ \ \ \mbox{(since $k$ is a DMC)} \nonumber \\
=      & \sum_{t = 1}^n I(I^n(t); O^n(t)) \nonumber \\
\leq & n I(T, k) \nonumber \\
         & \ \ \ \mbox{(by the convexity $I(\cdot, k)$)}
\end{align}
Thus,
\begin{align}
\inf_{k \in \mathcal A} I(I^n; O^n)  \leq \inf_{k \in \mathcal A} n I(T, k)
\end{align}
Now,
\begin{align}
\inf_{k \in \mathcal A} n I(T, k) \leq  \sup_{Q \in \mathcal P(\mathcal I)} \inf_{k \in \mathcal A} n I(Q, k) 
\end{align}
Thus, 
\begin{align}
\inf_{k \in \mathcal A} I(I^n; O^n)  \leq  \sup_{Q \in \mathcal P(\mathcal I)} \inf_{k \in \mathcal A} n I(Q, k) = nC^I(k \in \mathcal A)
\end{align}
This proves (\ref{AmosStep1SubStep2}). 

By previously stated arguments, it follows that the capacity of $k \in \mathcal A$ is $\geq R^E_X(D)$. 

Constant composition codes with distribution $T$ can be used for reliable communication over $k \in \mathcal A$ at rates $<R^E_X(D)$. $T$ is the input distribution to the input of $k \in \mathcal A$ in the original architecture for communication of i.i.d. $X$ source to within an expected distortion $D$ over $k \in \mathcal A$. Thus, the channel input distribution is unchanged and thus, the channel resource consumption in the architecture for reliable communication at rates $<R^E_X(D)$ over $k \in \mathcal A$ is the same as the channel resource consumption in the original architecture for communication of i.i.d. $X$ source to within distortion $D$ over $k \in \mathcal A$.

\section{Details of the calculations of bounds on probabilities of events (\ref{CorrectDecodingEvent}) and (\ref{ErrorEvent}) and the symmetry argument to prove (\ref{JJMainDualityJJ})} \label{CalculationAppendix}

Bound for probability of event (\ref{CorrectDecodingEvent}):
\begin{align}
       & \Pr \left ( \left \{ \frac{1}{n'} d^{n'}(U_c^{n'}, Y^{n'}) \leq D \right \} \right . \cap \nonumber \\
       & \hspace{2cm}  \left . \cap_{i = 1}^{2^{\lfloor n'R \rfloor} - 1} \left \{ \frac{1}{n'} d^{n'} ({U'}_i^{n'}, Y^{n'}) > D  \right \} \right ) \nonumber \\
=    & \Pr \left ( \left \{ \frac{1}{n'} d^{n'}(U_c^{n'}, Y^{n'}) \leq D \right \} \right ) + \nonumber \\
      &  \hspace{0.15cm} \Pr \left ( \cap_{i = 1}^{2^{\lfloor n'R \rfloor} - 1} \left \{ \frac{1}{n'} d^{n'} ({U'}_i^{n'}, Y^{n'}) > D  \right \}\right ) -  \nonumber \\
       &  \hspace{2.3cm} \Pr \left ( \left \{ \frac{1}{n'} d^{n'}(U_c^{n'}, Y^{n'}) \leq D \right \} \right . \cup \nonumber \\
       & \left . \hspace{2cm}  \cap_{i = 1}^{2^{\lfloor n'R \rfloor} - 1} \left \{ \frac{1}{n'} d^{n'} ({U'}_i^{n'}, Y^{n'}) > D  \right \} \right ) \nonumber \\
\geq & (1 - \omega_{n'}) + \nonumber \\ 
       & \hspace{.4cm} \Pr \left ( \cap_{i = 1}^{2^{\lfloor n'R \rfloor} - 1} \left \{ \frac{1}{n'} d^{n'} ({U'}_i^{n'}, Y^{n'}) > D  \right \}\right ) - \nonumber \\ 
       & \hspace{7.3cm} 1 \nonumber \\
=       & - \omega_{n'} + \Pr \left ( \cap_{i = 1}^{2^{\lfloor n'R \rfloor} - 1} \left \{ \frac{1}{n'} d^{n'} ({U'}_i^{n'}, Y^{n'}) > D  \right \}\right ) \nonumber \\
=       & - \omega_{n'} + \prod_{i=1}^{2^{\lfloor n'R \rfloor } - 1} \Pr \left ( \left \{ \frac{1}{n'} d^{n'} ({U'}_i^{n'}, Y^{n'}) > D \right \} \right ) \nonumber \\ 
          & \mbox{ (since ${U'}_i^{n'}, 1 \leq i \leq 2^{\lfloor n'R \rfloor } - 1$, $Y^{n'}$ are} \nonumber \\
          & \hspace{2.5cm} \mbox{ independent random variables)} \nonumber \\
=       & - \omega_{n'}  + \nonumber \\ 
         & \hspace{1cm} \left [ \Pr \left ( \left \{ \frac{1}{n'} d^{n'} ({U}^{n'}, Y^{n'}) > D \right \} \right )  \right ] ^ { 2^{\lfloor n'R \rfloor } - 1} \nonumber \\
         & \mbox{(where $U^{n'}$ is uniform on $\mathcal U^{n'}$} \nonumber \\
         & \hspace{3cm} \mbox{and is independent of $Y^{n'}$)} \nonumber \\
=       &  - \omega_{n'} +  \nonumber \\
         & \hspace{.5cm}   \left [   \sum_{y^{n'} \in \mathcal Y^{n'}} p_{Y^{n'}}(y^{n'}) \Pr \left ( \frac{1}{n'} d^{n'} ({U}^{n'}, Y^{n'})  \right . \right . \nonumber \\
        &\hspace{1.3cm}  \left . \textcolor{white}{\sum_{y^{n'} \in \mathcal Y^{n'}}} \left . \textcolor{white}{\frac{1}{n'}} > D \ \Bigg | \ Y^{n'} = y^{n'}  \right )      \right ] ^ { 2^{\lfloor n'R \rfloor } - 1} \nonumber \\
=       &  - \omega_{n'} +  \nonumber \\
       &  \hspace{.5cm}  \left [   \sum_{y^{n'} \in \mathcal Y^{n'}} p_{Y^{n'}}(y^{n'}) \Pr \left ( \frac{1}{n'} d^{n'} ({U}^{n'}, y^{n'})  \right . \right . \nonumber  \\
       & \hspace{1.3cm} \left . \textcolor{white}{\sum_{y^{n'} \in \mathcal Y^{n'}}} \left .  \textcolor{white}{\frac{1}{n'}} > D \ \Bigg | \ Y^{n'} = y^{n'}  \right )      \right ] ^ { 2^{\lfloor n'R \rfloor } - 1} \nonumber \\
=       &  - \omega_{n'} +  \nonumber \\
   & \hspace{.5cm}  \left [   \sum_{y^{n'} \in \mathcal Y^{n'}} p_{Y^{n'}}(y^{n'}) \Pr \left ( \frac{1}{n'} d^{n'} ({U}^{n'}, y^{n'}) \right . \right . \nonumber \\ 
   & \hspace{3.2cm} \left . \textcolor{white}{\sum_{y^{n'} \in \mathcal Y^{n'}}} \left . \textcolor{white}{\frac{1}{n'}}  > D  \right )      \right ] ^ { 2^{\lfloor n'R \rfloor } - 1} \nonumber \\
          & \hspace{2cm} \mbox{ (since $U^{n'}$ and $Y^{n'}$ are independent) } \nonumber \\ 
\geq    &  -\omega_{n'} +  \nonumber \\ 
         & \hspace{1cm} \left [ \inf_{y^{n'} \in \mathcal Y^{n'}} \Pr \left ( \left \{ \frac{1}{n'} d^{n'} ({U}^{n'}, y^{n'})  \right . \right . \right . \nonumber \\ 
       & \hspace{2.9cm} \left . \textcolor{white}{\inf_{y^{n'} \in \mathcal Y^{n'}}} \left .  \left .  \textcolor{white}{\frac{1}{n'}}  > D \right \} \right ) \right ] ^{ 2^{\lfloor n'R \rfloor } - 1} \nonumber \\
=          &  -\omega_{n'} +  \nonumber \\ 
         & \hspace{1cm} \left [ \inf_{q \in \mathcal G^{n'}} \Pr \left ( \left \{ \frac{1}{n'} d^{n'} ({U}^{n'}, y_q^{n'})  \right . \right . \right . \nonumber \\ 
       & \hspace{2.9cm} \left . \textcolor{white}{\inf_{q \in \mathcal G^{n'}}} \left .  \left .  \textcolor{white}{\frac{1}{n'}}  > D \right \} \right ) \right ] ^{ 2^{\lfloor n'R \rfloor } - 1}
\end{align}
The last equality above follows because
\begin{align}
\Pr \left ( \left \{ \frac{1}{n'} d^{n'} ({U}^{n'}, y^{n'}) > D \right \} \right )
\end{align}
depends only on the type of $y^{n'}$; see the symmetry argument later in the appendix. This gives the desired bound on probability of event (\ref{CorrectDecodingEvent}).

Bound for probability of event (\ref{ErrorEvent}):
\begin{align}
           & \Pr \left ( \cap_{i=1}^{2^{\lfloor n'R \rfloor}} \left \{  \frac{1}{n'}d^{n'}(u^{n'}, V_{q, i}^{n'}) > D \right \} \right )  =  \nonumber  \\
          &  \hspace{.6cm}  \prod_{i=1}^{2^{\lfloor n'R \rfloor}}  \Pr \left (  \left \{  \frac{1}{n'}d^{n'}(u^{n'}, V_{q, i}^{n'}) > D \right \} \right ) =  \nonumber \\ 
        & \hspace{1.2cm} \left [  \Pr \left (  \left \{  \frac{1}{n'}d^{n'}(u^{n'}, V_{q, i}^{n'}) > D \right \} \right )\right ] ^{2^{\lfloor n'R \rfloor }}
\end{align}
where $V_q^{n'}$ is uniform on $\mathcal V_q^{n'}$.

There is choice of $q \in \mathcal G^{n'}$. Thus, a bound for the probability of the event is
\begin{align}
 \left [  \inf_{q \in \mathcal G^{n'}} \Pr \left (  \left \{  \frac{1}{n'}d^{n'}(u^{n'}, V_q^{n'}) > D \right \} \right )\right ] ^{2^{\lfloor n'R \rfloor }}
\end{align}
This gives the desired bound on the probability of event (\ref{ErrorEvent}).

The symmetry argument to prove (\ref{JJMainDualityJJ}):

First step is to prove that 
\begin{align}\label{SymmetryStep1}
& \Pr \left ( \left \{ \frac{1}{n'} d^{n'} ({U}^{n'}, y_q^{n'}) > D \right \} \right ) =  \nonumber \\
& \hspace{3cm} \Pr \left (  \left \{ \frac{1}{n'} d^{n'} ({U}^{n'}, {y'_q}^{n'}) > D \right \} \right )
\end{align}
for sequences $y_q^{n'}$ and ${y'_q}^{n'}$ with type $q$. Since $U^{n'}$ is the uniform distribution on $\mathcal U^{n'}$, it follows that it is sufficient to prove that the sets
\begin{align}
& \left \{ u^{n'} : \frac{1}{n'} d^{n'} ({u}^{n'}, y_q^{n'}) > D \right \} \ \mbox{and} \nonumber \\ 
& \hspace{3cm}  \left \{ u^{n'} : \frac{1}{n'} d^{n'} ({u}^{n'}, {y'_q}^{n'}) > D \right \}
\end{align}
have the same cardinality. ${y'_q}^{n'} = \pi^{n'} y_q^{n'}$ for some permutation $\pi^{n'}$ since ${y'_q}^{n'}$ and $y_q^{n'}$ have the same type. Denote the sets
\begin{align}
\mathcal B_{y_q^{n'}} \triangleq  \left \{ u^{n'} : \frac{1}{n'} d^{n'} ({u}^{n'}, y_q^{n'}) > D \right \}
\end{align}
Set $\mathcal B_{{y'_q}^{n'}}$ is defined analogously.

Let $u^{n'} \in \mathcal B_{y_q^{n'}}$. Since the distortion function is permutation invariant, $d^{n'}(\pi^{n'}u^{n'},\pi^{n'} y_q^{n'})$   $=$ $d^{n'}(u^{n'}, y_q^{n'})$. Thus, $\pi^{n'}u^{n'} \in \mathcal B_{{y'_q}^{n'}}$. If $u^{n'} \neq u'^{n'}$, $\pi^{n'}u^{n'} \neq \pi^{n'}u'^{n'}$. It follows that $|\mathcal B_{{y'_q}^{n'}}| \geq |\mathcal B_{y_q^{n'}}|$. Interchanging $y_q^{n'}$ and ${y'_q}^{n'}$ in the above argument, $|\mathcal B_{y_q^{n'}}| \geq |\mathcal B_{{y'_q}^{n'}}|$. It follows that $|\mathcal B_{{y_q}^{n'}}| = |\mathcal B_{{y'_q}^{n'}}|$. (\ref{SymmetryStep1}) follows.

Let $V_q^{n}$ be independent of $U^{n'}$. From  (\ref{SymmetryStep1}) it follows that 
\begin{align}\label{SymmetryStep1Conclusion}
& \Pr \left ( \left \{ \frac{1}{n'} d^{n'} ({U}^{n'}, y^{n'}) > D \right \} \right )  = \nonumber \\ 
& \hspace{3cm} \Pr \left ( \left \{ \frac{1}{n'} d^{n'} ({U}^{n'}, V_q^{n'}) > D \right \} \right ) 
\end{align}
By an argument identical with the one used to prove (\ref{SymmetryStep1}), it follows that 
\begin{align} \label{SymmetryStep2}
& \Pr \left ( \left \{ \frac{1}{n'} d^{n'} ({u}^{n'}, V_q^{n'}) > D \right \} \right ) = \nonumber \\
& \hspace{3cm}  \Pr \left ( \left \{ \frac{1}{n'} d^{n'} ({u'}^{n'}, V_q^{n'}) > D \right \} \right )
\end{align}
for $u^{n'}, u'^{n'} \in \mathcal U^{n'}$. From (\ref{SymmetryStep2}) it follows that 
\begin{align} \label{SymmetryStep2Conclusion}
& \Pr \left ( \left \{ \frac{1}{n'} d^{n'} ({u}^{n'}, V_q^{n'}) > D \right \} \right ) = \nonumber \\ 
& \hspace{3cm} \Pr \left ( \left \{ \frac{1}{n'} d^{n'} ({U}^{n'}, V_q^{n'}) > D \right \} \right )
\end{align}
From (\ref{SymmetryStep1Conclusion}) and (\ref{SymmetryStep2Conclusion}), (\ref{JJMainDualityJJ}) follows.
\end{document}